\documentclass[prd,,preprintnumbers,nofootinbib,twocolumn]{revtex4-1}
\usepackage{graphicx}
\usepackage{amsmath}
\usepackage{amssymb}
\usepackage[bookmarksopen,bookmarksnumbered]{hyperref}
\usepackage{color}

\let\oldleft\left
\def\xleft{\mathopen{}\oldleft}

\newcommand\3[1]{\boldsymbol{#1}}

\newcommand{\MSbar}{\ensuremath{ \overline{\rm MS} }}

\newcommand{\VC}[1]{%
  \begin{tabular}[c]{l}%
    #1%
  \end{tabular}
}

\begin{document}


\title{Equality of Two Definitions for Transverse Momentum
  Dependent Parton Distribution Functions}
  
\preprint{YITP-SB-12-35}  

\author{John Collins}
\email{collins@phys.psu.edu}
\affiliation{%
  104 Davey Lab, Penn State University, University Park PA 16802, USA
}
\author{Ted Rogers}
\email{rogers@insti.physics.sunysb.edu}
\affiliation{%
  C. N. Yang Institute for Theoretical Physics, Stony Brook
  University, Stony Brook, NY 11794-3840, USA 
}


\date{8 November 2012}

\begin{abstract}
    We compare recent, seemingly different, approaches to 
    TMD-factorization
    (due to Echevarria, Idilbi, and Scimemi and to Collins),
    and show that they are the same,
    apart from an apparent difference in their definition of the
    \MSbar{} renormalization scheme.
\end{abstract}

\maketitle


\section{Introduction}

In a number of processes, notably the Drell-Yan process at small
transverse momentum, it is important to use
transverse-momentum-dependent (TMD) parton distribution functions
(pdfs) and an associated TMD factorization theorem of QCD.  However,
there is a variety of different formulations in the literature, with
seemingly very different methodologies and different definitions of
the TMD pdfs.

Some of the differences even amount to apparent incompatibilities.
For example, in the formalism developed by Collins, Soper, and Sterman
(CSS) \cite{Collins:1981uw,Collins:1984kg}, TMD pdfs depend not only
on the essential kinematic parameters $x$ and $k_T$, and on a
renormalization scale $\mu$, but also on an extra auxiliary scale $\zeta$.
An essential role is played by the Collins-Soper (CS) evolution
equation \cite{Collins:1981uw} of the pdfs with respect to $\zeta$.  In
contrast, in the recent work by Echevarria, Idilbi, and Scimemi (EIS)
\cite{GarciaEchevarria:2011rb} in the context of soft-collinear
effective theory (SCET), there appears to be no corresponding
parameter.  Indeed, in a phenomenological application of this work by
Echevarria, Idilbi, Sch\"afer, and Scimemi \cite{Echevarria:2012pw},
there even appears the statement that they ``do not have to solve
Collins-Soper equation''.  Furthermore, in the recent improvement of
the CSS method by one of us (JCC) in Ref.~\cite{Collins:2011qcdbook},
the presence of the $\zeta$ parameter appears to be intimately associated
with the use of non-light-like Wilson lines in JCC's gauge-invariant
definition of the TMD pdfs. But all the Wilson lines in EIS's
definition are light-like (with the aid of their $\delta$-regulator in a
limit that the regulator is removed).  Finally, the definitions in the
two methods appear very different: In the JCC method the finite TMD
pdf is \cite[Eq.\ (13.106)]{Collins:2011qcdbook} a renormalized
convolution product of a bare TMD pdf with a square root of three bare
soft factors, two of them with the characteristic non-light-like
Wilson lines. But in the EIS definition \cite[Eq.\
(2.13)]{GarciaEchevarria:2011rb}, the square root involves only one
bare soft factor.  JCC and EIS do agree on the presence of the square
root.

In this paper, we show that nevertheless the EIS and JCC definitions
give exactly the same TMD pdfs (and hence the same hard scattering
factors in TMD factorization properties).  We show how the EIS
definitions do in fact have a parameter $\zeta$ that EIS have identified
with $Q^2$.  Thus their TMD pdfs obey a CS equation.  Although the
application of the EIS work in \cite{Echevarria:2012pw} does not
explicitly use the CS equation, it actually uses equivalent results,
the most notable of which are the results of Korchemsky and Radyushkin
\cite{Korchemsky:1987wg} on the $Q$-dependence of the anomalous
dimension of a cusp anomalous dimension.  We show how the JCC
definition of the TMD pdfs can be cast into exactly the form of the
EIS definition, in which no non-light-like Wilson lines appear and in
which there is only a single bare soft factor instead of three.  It
should be noted that, in both cases, the bare factors are separately
divergent because of the presence of light-like Wilson lines, which
must be regulated, and a limit taken with the regulator removed.  The
EIS and JCC differ in the preferred method to regulate light-like
Wilson lines.  The EIS form of definition requires coordination of
the rates at which are removed the regulators of different light-like
Wilson lines.

EIS do say in \cite{Echevarria:2012pw} that ``one can argue that the
two [methods] are equivalent'', but they do not provide any details of
the argument.\footnote{See also \cite{Echevarria:2012qe}, which
  appeared as we were completing this paper.}

We expect that the methods developed in this paper can be applied to
analyze and/or relate many other methods for formulating TMD
factorization.  A list of some relevant papers beyond those cited
above is \cite{Ji:2004wu, Ji:2004xq, Cherednikov:2007tw,
  Cherednikov:2009wk, Mantry:2010bi, Becher:2010tm, Chay:2012mh}.

The reader needs to be very aware that 
the JCC method we use
here differs in very important details from the earlier CS formalism
on which it is based, and also that there was a change in candidate
definitions of collinear factors in the course of the derivations in
\cite{Collins:2011qcdbook}.  The definitions that we use are of the
form given in \cite[Sec.\ 10.11]{Collins:2011qcdbook} for the Sudakov
form factor, and in \cite[Sec.\ 13.15.3]{Collins:2011qcdbook} for TMD
pdfs (with a modification to use past-pointing Wilson lines for the
Drell-Yan process, as summarized at the beginning of \cite[Sec.\
14.5]{Collins:2011qcdbook}).

The paper is structured as follows:  In Sect.~\ref{sec:common} we introduce
notation and enumerate the common features of the EIS and JCC methods.
In Sect.~\ref{sec:finiteness}, we prove that the EIS and JCC 
methods both give the same result for the cross section.
We prove in Sect.~\ref{sec:JCCformula} that the JCC and EIS TMD PDFs are 
equal.  In Sect.~\ref{sec:EISzeta}, we identify the parameter in the EIS TMD 
PDFs that corresponds to $\zeta$ in the JCC method.  
We summarize our results in Sect.~\ref{sec:summary}.

\section{Common Overall Structure}
\label{sec:common}

We work with the Drell-Yan process: production of a high-mass lepton
pair of 4-momentum $q$ from a single virtual electroweak boson in a
collision of hadrons of momenta $p_A$ and $p_B$.  The invariant mass
of the lepton pair is $Q=\sqrt{q^\mu q_\mu}$, its rapidity is $y$, and its
transverse momentum is $q_T$. 

When $q_T \ll Q$, TMD factorization as formulated by both EIS and JCC is
that the cross section is proportional to a Fourier transform in
transverse coordinate space of TMD pdfs:
\begin{multline}
  \label{eq:fact.1}
  \frac{d\sigma}{d^4q}
  =
  \sum_{i,j} \int d^2\3{b}_T e^{i\3{b}_T\cdot\3{q}_T}
\\
  \tilde{f}_{i/A}\xleft( x_A, b_T \right) \tilde{f}_{i/B}\xleft( x_B, b_T \right) H_{ij}(Q)
  + \mbox{p.s.c.}
\end{multline}
Here, the sum is over the relevant parton flavors, $\tilde{f}_{i/A}$
and $\tilde{f}_{j/B}$ are Fourier transforms into transverse
coordinate space of TMD PDFs, $H_{ij}$ is a hard scattering, and
``p.s.c.''  denotes ``power-suppressed corrections''.  The corrections
to the formula are either suppressed by a power of $q_T/Q$ or by a
power of $M/Q$, where $M$ denotes a typical scale for light hadronic
masses; those corrections suppressed only by a power of $q_T/Q$ can be
handled by normal factorization methods with integrated pdfs, as by
the $Y$ term in the CSS formalism.

To simplify the notation, we write the above structure as
\begin{equation}
  \label{eq:fact}
  \sigma = A B H,
\end{equation}
with all the kinematic variables, flavor indices, and corrections left
implicit.  The intuitive interpretation is that the factors $A$ and
$B$ contain dependence on momenta collinear to each of the
corresponding beam particles.

As for the kinematics, we have $x_Ax_Bs = Q^2$ and the rapidity of the
Drell-Yan pair in the center-of-mass frame of the beams is
$y=\frac12\ln\frac{x_A}{x_B}$. 
In light-front coordinates, the beam momenta are
\begin{equation}
  \label{eq:pA.pB}
  p_A = \left( p_A^+, \frac{m_A^2}{2p_A^+}, 0_T \right),
\qquad
  p_B = \left( \frac{m_B^2}{2p_B^-}, p_B^-, 0_T \right),
\end{equation}
with $p_A^+$ and $p_B^-$ large, so that $s\simeq 2p_A^+p_B^-$.
The plus momentum of the parton from $A$ is $k_A^+=x_Ap_A^+$,
and the minus momentum of the parton from $B$ is $k_B^-=x_Bp_B^-$.

The above statement of TMD factorization is used by both the JCC and
EIS methods.  However, in \eqref{eq:fact.1} we have indicated only the
dependence on kinematic parameters of the factors $\tilde{f}_{i/A}$
$\tilde{f}_{j/B}$, and $H_{ij}$.  In addition all three factors depend
on auxiliary parameters.  It is common to both methods that all the
factors depend on the renormalization scale $\mu$.  There is one further
parameter for the TMD pdfs in both methods.

In the JCC approach, each collinear factor also depends on a $\zeta$
parameter.  This is constructed from an arbitrarily chosen rapidity
parameter $y_n$, which has the intuitive purpose of separating right
and left moving quanta.  For $A$, the $\zeta$ parameter is
\begin{equation}
  \label{eq:zeta.A}
  \zeta_A = 2(k_A^+)^2 e^{-2y_n} = 2(x_Ap_A^+)^2 e^{-2y_n} ,
\end{equation}
and for $B$ it is
\begin{equation}
  \label{eq:zeta.B}
  \zeta_B = 2(k_B^-)^2 e^{2y_n} = 2(x_Bp_B^-)^2 e^{2y_n} .
\end{equation}
These two quantities obey $\zeta_A\zeta_B = Q^4$.  In both the EIS and JCC approaches, 
neither the hard scattering $H$ nor the cross section has dependence 
on the $\zeta$ variables.

In the case of the EIS method, the authors' statements and notation suggest that
the TMD PDFs depend on $\mu$ but not on some other parameter like $\zeta$
--- see the paragraph above \cite[Eq.\
(1.6)]{GarciaEchevarria:2011rb}, and see \cite[Eq.\
(2.15)]{GarciaEchevarria:2011rb}. In fact, the EIS PDFs do 
depend on $Q$ as a separate parameter distinct from $\mu$.  This can be seen from
\cite[Eq.\ (2.17)]{GarciaEchevarria:2011rb}, where there is $Q$ and
$\mu$ dependence for the coefficient $C_n$ that relates the TMD PDF to
the ordinary integrated PDF.  It can also be seen from the fact that
the anomalous dimension of their TMD PDF depends on both $Q$ and $\mu$
--- see \cite[Eq.\ (3.21)]{GarciaEchevarria:2011rb}. 
In later sections of this paper, we will show how the $Q^2$ parameter
of the TMD pdfs with the EIS definition should in fact be identified
with the $\zeta$ parameter of those with the JCC definition.

Common to both schemes is the property that the product of the two
TMD functions is related to a combination of unsubtracted TMD
functions and a soft function:
\begin{equation}
\label{eq:AB.0}
  AB = \frac{A_0B_0}{S_0} \times \mbox{UV renormalization}.
\end{equation}
Note that all the authors agree on applying UV renormalization in the
\MSbar{} scheme to remove divergences where transverse momenta go to
infinity \footnote{But see Sec.\ \ref{sec:MSbar} for a possible
  difference in the definitions of the \MSbar{} scheme.}.
Such divergences are all regulated by dimensional
regularization. 

Each of the unsubtracted functions is defined with the aid of
light-like Wilson lines, 
taken as a limit \footnote{
  Note that these Wilson lines must be joined by transverse links at
  infinity to give exactly gauge-invariant quantities
  \cite{Belitsky:2002sm,Ji:2002aa}.  Only with these transverse links
  can correct results be obtained in all gauges.  Notably, the
  transverse links are essential to get the correct Sivers asymmetry
  in light-cone gauge \cite{Belitsky:2002sm,Ji:2002aa}.  See
  Refs.~\cite{GarciaEchevarria:2011md,Idilbi:2004vb} for corresponding
  statements in SCET.
}.  
The $A_0$ and $B_0$ are the ``unsubtracted'' TMD pdfs $\tilde{f}^{\rm
  unsub}$ in~\cite[Eq.\ (13.106)]{Collins:2011qcdbook}, or the $\hat{f}_n$ 
functions in~\cite[Eq.\ (3.21)]{GarciaEchevarria:2011rb}.
The $S_0$ is a light-cone regulated soft function, defined according 
to~\cite[Eq.\ (13.106)]{Collins:2011qcdbook}
and~\cite[Eq.\ (2.7)]{GarciaEchevarria:2011rb}.
The division by $S_0$ cancels double
counting of soft gluon contributions between the $A_0$ and $B_0$
factors and also cancels the rapidity divergences that would otherwise
occur because light-like Wilson lines are used in $A_0$ and $B_0$.
Since each of the factors in \eqref{eq:AB.0} separately contain 
light cone divergences, a meaning can only be given by applying regulators 
to the Wilson lines, and then taking the limit of removing the regulators in a 
manner consistent with factorization.

As explained in \cite[Sec.\ 10.11.1]{Collins:2011qcdbook}, the limit
of light-like Wilson lines and the removal of dimensional
regularization do \emph{not} commute.  We will always define that the
limit light-like Wilson lines is to be applied first.  Then \MSbar{}
renormalization is applied and dimensional regularization removed.
EIS's work appears to use this order as well. 

\begin{figure}
  \centering
  \includegraphics[scale=.5]{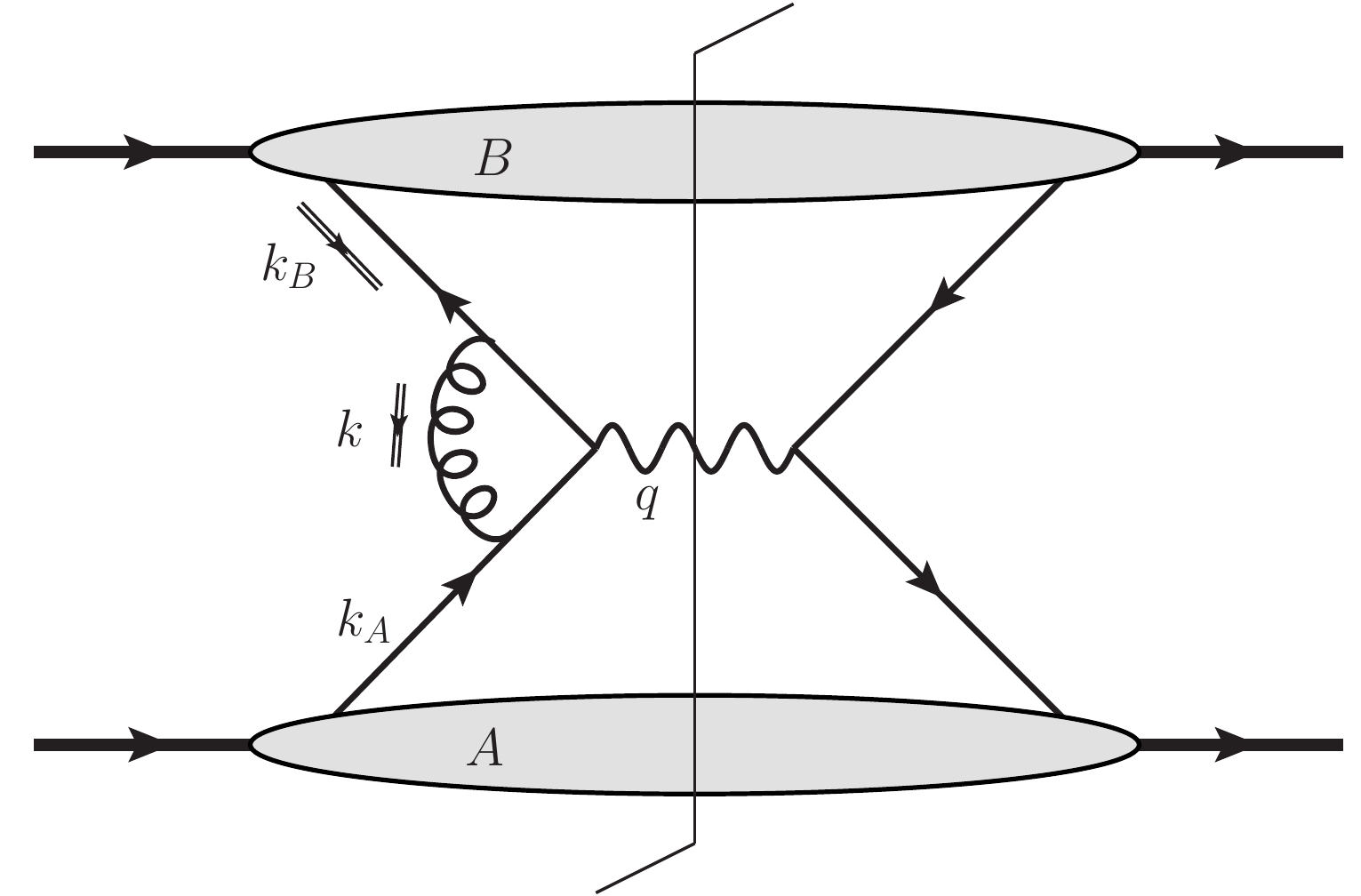}
  \caption{One-loop graph contributing to the Drell-Yan process.}
  \label{fig:DY.1loop}
\end{figure}

At this point, the only relevant difference between the EIS and JCC 
treatments is the choice of method for regulating the divergences that 
come from using light-like Wilson lines.  The JCC method uses non-light-like 
Wilson lines, and takes the limit that they becomes light-like:
\begin{multline}
\label{eq:AB.0.JCC}
  A_{\text{JCC}} B_{\text{JCC}}
  = \lim_{{y_1\to+\infty \atop y_2\to-\infty}}
        \frac{ A_{\text{JCC},0}\xleft( y_2 \right)
               B_{\text{JCC},0}\xleft( y_1 \right) }
             { S_{\text{JCC},0}\xleft( y_1,y_2 \right) }
\\
  \times \mbox{UV renormalization}.
\end{multline}
(This result follows directly from \cite[Eqs.\ (13.106) and
(14.31)]{Collins:2011qcdbook}.) 
The Wilson lines appear in Feynman diagram calculations 
as eikonal propagators of the form,
\begin{equation}
\label{eq:jcceikonal}
  \frac{i}{-k^+ + e^{2 y_2} k^-  + i0}, 
\qquad
  \frac{i}{- e^{-2 y_1} k^+ + k^- + i 0},
\end{equation}
together with the appropriate vertex factors.  Here, $y_1$ and $y_2$
are the finite Wilson line rapidities, and the signs are those
appropriate for when the gluon momenta are oriented as in the one-loop
graph for the Drell-Yan process in Fig.\ \ref{fig:DY.1loop}.
Note that, in the JCC method the regulated factors are all
matrix elements of exactly gauge-invariant operators.

The EIS approach \cite{GarciaEchevarria:2011rb} uses what they call a
$\delta$-regulator instead of \eqref{eq:jcceikonal}, where the Wilson line 
propagators for regulated light-like lines have the following forms:
\begin{equation}
  \label{eq:delta.def}
  \frac{i}{ -k^+ + i\delta^+ },
\qquad
  \frac{i}{ k^- + i\delta^- }.
\end{equation}
Their equivalent of \eqref{eq:AB.0.JCC} is
\begin{multline}
\label{eq:AB.0.EIS}
  A_{\text{EIS}} B_{\text{EIS}}
  = \lim_{\delta^+, \delta^- \to 0}
        \frac{ A_{\text{EIS},0}\xleft( \delta^+ \right) 
               B_{\text{EIS},0}\xleft( \delta^- \right) }
             { S_{\text{EIS},0}\xleft( \delta^-,\delta^+ \right) }
\\
  \times \mbox{UV renormalization}.
\end{multline}
That is, instead of using Eq.~\eqref{eq:jcceikonal}, they replace 
light-like propagators with ones with fixed non-zero auxiliary terms inserted 
into the denominators.

The right-hand-sides Eqs.~(\ref{eq:AB.0.JCC}) and (\ref{eq:AB.0.EIS})
differ only in how the light-like Wilson lines are regulated, and
therefore the limit as the regulator is removed must be the same in
each case.  The same UV renormalization prescription, \MSbar{}, is
used.  Hence the product of the two TMD pdfs on the left-hand-sides is
the same, i.e., $AB$ is the same in the EIS and JCC schemes.  The
remaining issue is to determine whether the individual TMD pdfs, $A$
and $B$, are the same in the two schemes.

Our proof that this is indeed the case will use factorization
properties for the unsubtracted factors, and one immediate consequence
of these factorization properties is that the limits in
Eqs.~(\ref{eq:AB.0.JCC}) and (\ref{eq:AB.0.EIS}) do exist. 

One complication is that EIS also use a corresponding regulator for
the quark lines in their 
calculations, for example, \cite[Sec.\ 3]{GarciaEchevarria:2011rb}.
This provides a regulator for the collinear divergences associated
with quark-gluon interactions in massless QCD.  The regulators for the
quark lines and the Wilson lines have to correspond: Their regulator
parameters of the quark lines are denoted $\Delta^+$ and $\Delta^-$,
and they set $\Delta^+ = \delta^+ p_B^-$ and $\Delta^- = \delta^-
p_A^+$ (in our notation).  This coherence is needed so that the
combination $A_0B_0/S_0$ is a valid approximation to the original
graphs for the cross section.

However, for defining the parton densities, and for using
\eqref{eq:AB.0.EIS}, one can ignore this issue, provided that we also
use dimensional regulation.  Since dimensional regularization
regulates the collinear and soft divergences, we can take the limits
$\Delta^\pm\to0$ first.  Then the limits $\delta^\pm \to 0$ can be taken separately to
define the parton densities.  These results are compatible with the
factorization results obtained below.  However, as we will
demonstrate, the limits on $\delta^+$ and on $\delta^-$ need to be coordinated.

\section{Factorization Preserves  Finiteness and Regulator Independence of
 \texorpdfstring{$\lim \left( A_0B_0/S_0 \right)$}{lim(A0B0/S0)}}
 \label{sec:finiteness}

To see explicitly that the limits in \eqref{eq:AB.0.JCC} and \eqref{eq:AB.0.EIS}
exist and are the same, we apply factorization by the JCC method to
express the regulated $A_0$, $B_0$ and $S_0$ in terms of products of
the finite collinear factors in the JCC scheme.  Factorization is
applied separately for $A_0$ etc regulated by each of the JCC and EIS
methods.  (EIS would undoubtedly apply their factorization method
instead.)

In all of this paper, we will take for granted that factorization
holds for the cross section and for the other situations that we
consider.  Our aim is to use factorization to relate different
formulations.

There are a couple of complications that affect the organization of
our proofs:
\begin{itemize}
\item In constructing the TMD pdfs there are limiting operations: to
  remove the regulator(s) on light-like Wilson lines, and to remove
  dimensional regularization.  These limits do not commute --- see,
  e.g., \cite[Sec.\ 10.8]{Collins:2011qcdbook}.  The definitions are
  first to remove the regulator of light-like Wilson lines, e.g., in
  Eqs.~\eqref{eq:AB.0.JCC} and \eqref{eq:AB.0.EIS}, then to apply UV
  renormalization, and finally to remove dimensional regularization.
\item Corresponding to the different steps in taking limits, there are
  actually three different objects that implement some kind of TMD
  pdf:
  \begin{enumerate}
  \item The unsubtracted quantities $A_0$, etc, as on the right-hand
    side of Eqs.~\eqref{eq:AB.0.JCC} and \eqref{eq:AB.0.EIS}; to make
    these finite a regulator must be used on their light-like Wilson
    lines.  Were all divergences absent, these would give the natural
    definitions of TMD pdfs.
  \item The unsubtracted TMD pdfs are combined with square roots of
    unsubtracted soft factors, in definitions that we present below,
    in which the limit is taken that the regulator(s) of the
    light-like Wilson lines is removed.  (The JCC definition continues
    to have certain other non-light-like Wilson lines.)  These we will
    call ``unrenormalized'' TMD pdfs, notated $A_{\text{unren}}$, etc,
    and we will prove these are the same in the JCC and EIS
    definitions.
  \item Finally UV renormalization is applied and then dimensional
    regularization removed.  The resulting TMD pdfs are those on the
    left-hand side of Eqs.~\eqref{eq:AB.0.JCC} and
    \eqref{eq:AB.0.EIS}, they are finite in QCD, and they are the TMD
    pdfs that are directly used in phenomenological applications.
  \end{enumerate}
\end{itemize}

There are several kinds of divergence that appear at various stages of
using factorization.

First are the UV divergences in QCD itself; these are removed by
renormalization as part of the definition of QCD.

There are also UV divergences in the definitions of pdfs (and other
related quantities).  Although these divergences are also removed by
essentially conventional renormalization, their status is rather
different from those of QCD itself.  The definitions of pdfs arise
from the application of approximations appropriate to a collinear
region of momentum, and the UV divergences arise from a natural
extrapolation of the approximations to all momenta.  We will therefore
characterize these divergences as ``induced divergences''.  They
cancel between the factors in a factorization property (as do the
corresponding renormalization factors).  Induced UV divergences also
arise for the same reasons in the ordinary short-distance operator
product expansion (OPE).  It is possible to simply cut-off induced
divergences.  But a cleaner formalism results if one allows the
divergences to occur and defines finite operator matrix elements by
renormalization. 

Then there are the IR and collinear divergences that appear in
Feynman-graph calculations with massless quarks and gluons.  These
would be absent if all the fields have non-zero mass, as can be true
in a model theory.  They are presumably cutoff by (non-perturbative)
confinement in QCD.  In perturbatively calculable short-distance
quantities, like hard-scattering coefficients, it is useful to neglect
masses to simplify calculations.  But the IR and collinear divergences
must cancel in such quantities, precisely because they are
short-distance quantities.

Finally, there are rapidity divergences.  These arise when TMD pdfs 
are defined in terms of operator matrix elements with exactly light-like 
Wilson lines.  In a gauge theory such Wilson lines are a natural
consequence of the expansions in small variables for collinear and
soft momenta in setting up factorization; for this reason, they occur
in both the EIS and JCC methods.  
Having separately defined operator matrix elements is 
important for analyzing asymptotic behavior in the form of factorization
properties, and for identifying the operator structure of universal non-perturbative
objects like pdfs and fragmentation functions.  
These matrix elements can be studied with non-perturbative 
methods, and are of intrinsic interest as objects 
related to hadron structure.
Rapidity divergences are another example of induced divergences,
i.e., divergences introduced by the approximations used to
factorize the cross section.
To define finite pdfs, it is possible to cut off the rapidity
divergences in some fashion (as in the original CS formalism), but it
is also possible to remove them by a kind of generalization of
the renormalization that is applied to UV divergences.  Both the EIS
and JCC formalisms use such a generalized renormalization, implemented
by multiplying unsubtracted TMD pdfs by suitable square roots of
unsubtracted soft factors (or by some equivalent of this procedure).

Induced divergences, no matter whether they are UV divergences or
rapidity divergences, are unphysical in the sense that they do not
correspond to regions of momenta that give actual divergences in
unapproximated graphs for scattering.  Nevertheless they are symptoms
of important physical phenomena.  They arise when in some region of
momentum one expands to low order in powers of a small variable
relative to a much large variable.  Such expansions lead to the
definition of quantities like pdfs.  The natural extrapolation of the
expansion to all values of the expansion parameter gives the induced
divergences.  The range of momenta over which the expansion gives a
valid approximation to the original graph depends on the energy and
other kinematic parameters of the process under consideration.

Thus while induced divergences, including rapidity divergences, are in
one sense unphysical, they are also physical in another sense that
they are related (by extrapolation) to important, and even dominant,
physical effects.

Associated with the generalized renormalization of induced divergences
to construct finite pdfs, etc, are typically auxiliary scales like the
renormalization scale $\mu$ of \MSbar{} renormalization.  Such a scale
can be intuitively characterized as the range over which the
underlying approximation is used.  Since the range over which the
approximation is useful is process-dependent, evolution (e.g., by the
renormalization group) with respect to the auxiliary parameter(s)
quantifies how the actual cross section is related to particular
ranges of momenta inside the amplitudes for the process.

The generalized renormalization of rapidity divergences in the JCC
scheme leads to the auxiliary parameter $\zeta$ that we have already
mentioned, with the associated CS evolution equation.  As we will see,
the EIS method also uses corresponding properties.

\subsection{\texorpdfstring{$A_0$}{A0}, \texorpdfstring{$B_0$}{B0} 
and \texorpdfstring{$S_0$}{S0} in the JCC method} 

\begin{figure*}
  \centering
  \begin{tabular}{c@{\hspace*{10mm}}c}
       \VC{\includegraphics[scale=.4]{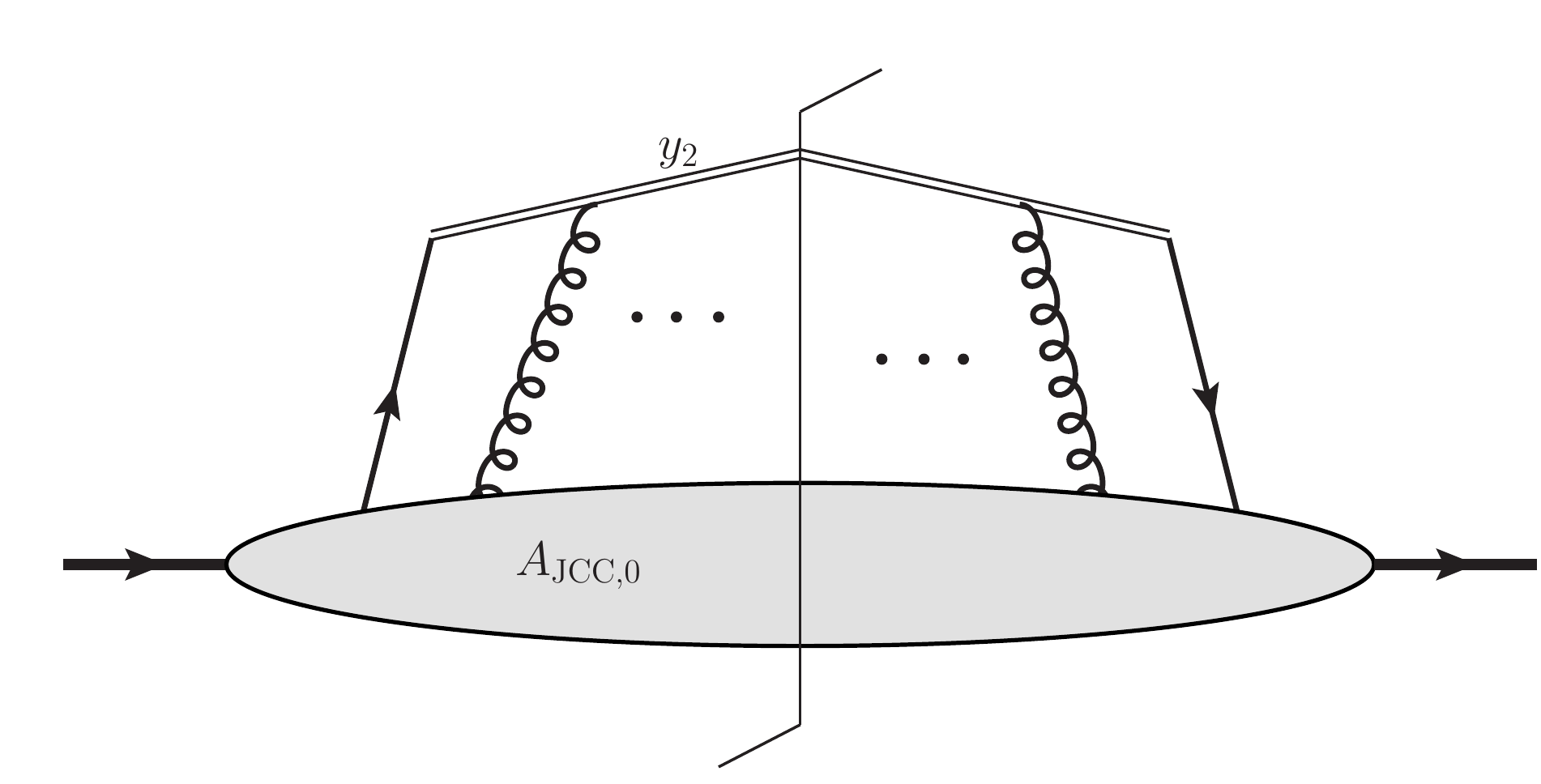}}
   & 
       \VC{\includegraphics[scale=.4]{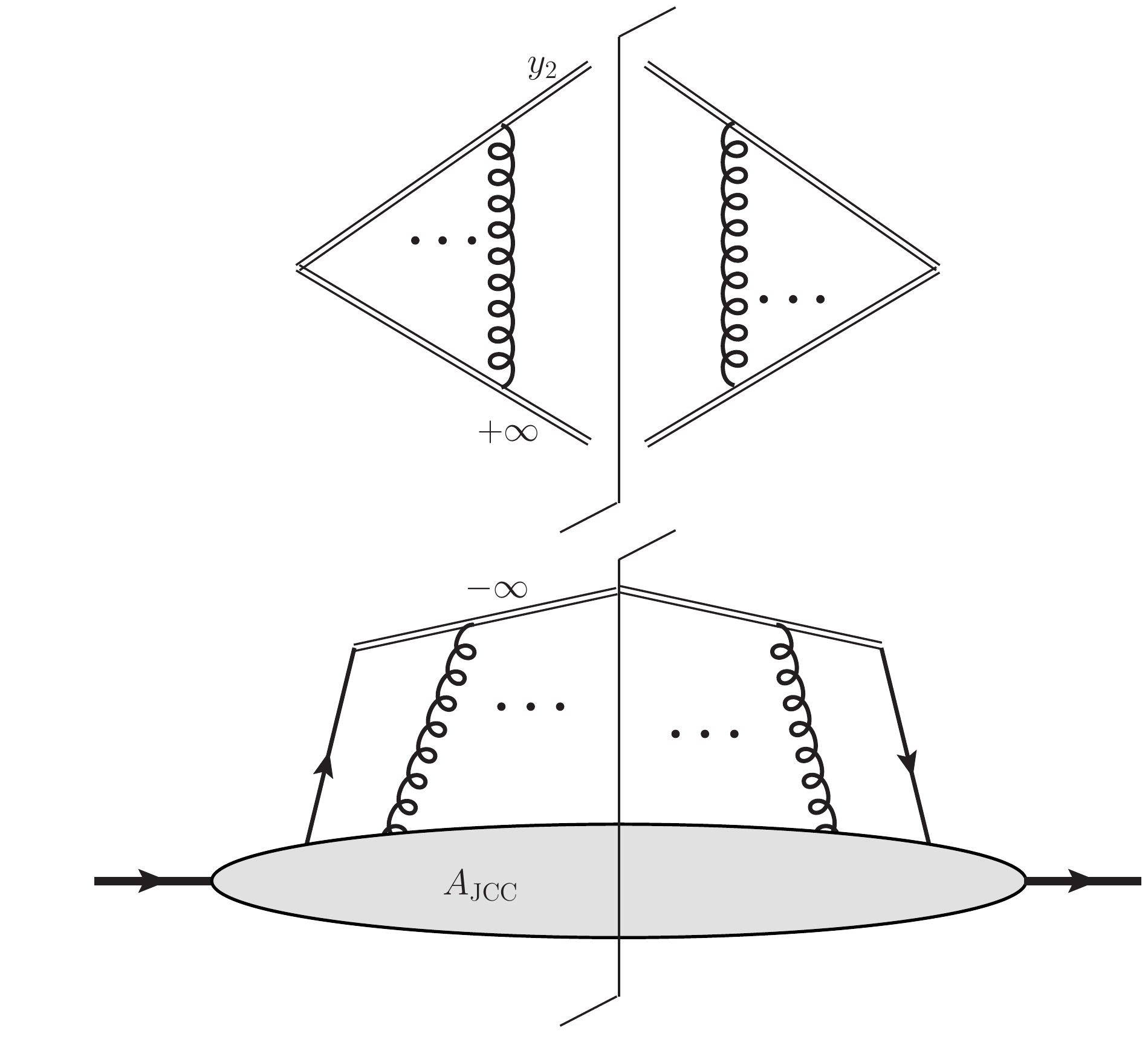}}
  \\
     (a) & (b)
  \end{tabular}
  \caption{For the factorization in \eqref{eq:A0.JCC.fact}:
     (a) the left-hand-side, and (b) the \emph{bare} collinear factors
     for the right-hand-side.
     Note that the two factors in (b) also have generalized
     renormalization by square roots of soft factors that are
     \emph{not} shown here.
  }
  \label{fig:A0.fact}
\end{figure*}

In $A_{\text{JCC},0}$, the hadron and the Wilson line of rapidity
$y_2$ are widely separated in rapidity, so they factor into two
collinear pieces:
\begin{multline}
\label{eq:A0.JCC.fact}
  A_{\text{JCC},0}\xleft( y_2 \right)
  = A_{\text{JCC,~unren.}}\xleft( 2(k_A^+)^2e^{-2y_n} \right)
\\
    \, C_{\text{JCC,~unren},W}\xleft( y_n-y_2 \right)
    + \mbox{error}.
\end{multline}
The factors are illustrated in Fig.~\ref{fig:A0.fact}.
As usual, the leading regions have collinear parts, a soft part, and a
hard part.  The collinear parts are associated with the hadron 
and with the Wilson line of rapidity $y_2$, and correspond to the factors
on the right.  No separate factor is needed for the soft part,
according to the methods of both JCC and EIS.  As usual in the JCC
scheme, the collinear factors depend on a rapidity parameter $y_n$,
which we choose to be the same as in \eqref{eq:zeta.A}, although our
argument can be generalized to deal with a different value.  Notice
that we indicate the dependence on the auxiliary parameters and the
Wilson line rapidities and not the other kinematic parameters (e.g.,
$x$ and $b_T$.)

As for the hard part, it is in fact to be replaced by its lowest order
value, which is unity.  The reason is that at this stage, we keep
dimensional regularization as we take $y_2\to-\infty$.  Now the hard scale is
set by the total energy of the process, which we can parameterize by
$k_A^+e^{-y_2}$.  With dimensional regularization in $4-2\epsilon$
dimensions, higher orders in the hard scattering are suppressed by a
power $1/(k_A^+e^{-y_2})^{\text{constant}\times\epsilon}$, and therefore vanish
when $y_2\to-\infty$ at fixed $\epsilon$.  (If we removed dimensional regularization
and renormalized everything before taking $y_2\to-\infty$, we would get
differently organized results; this is the non-commutativity of limits
mentioned earlier.)  Since we have not yet removed dimensional
regularization, we do not yet need to apply UV renormalization.
Consequently, we are working with the unrenormalized collinear
factors, as indicated by the notation on the right-hand-side of
Eq.~\eqref{eq:A0.JCC.fact}.

Apart from the issue of UV renormalization, the collinear factor
$A_{\text{JCC,~unren.}}$ is the same as in the JCC version of
Drell-Yan factorization.  That is, it is the TMD pdf associated with
hadron $A$, defined as in \cite[Sec.\
13.15.3]{Collins:2011qcdbook}, but with past-pointing Wilson lines, as
appropriate for the Drell-Yan process that we work with here.  It
includes factors involving the square roots of soft factors, and in
two of these factors, there is a Wilson line with a finite rapidity,
$y_n$.  The quantity $C_{\text{JCC},W}$ is a corresponding 
collinear
factor
associated with the Wilson line of rapidity $y_2$ in the
left-hand-side.

\begin{widetext}
We recall the definition (\cite[Eq.\ (13.106)]{Collins:2011qcdbook}) of $A_{\text{JCC}}$:
\begin{align}
\label{eq:A.JCC.def.copy}
  A_{\text{JCC,~unren}}\xleft( \zeta_A \right)
   & = \lim_{{y'_1\to+\infty \atop y'_2\to-\infty}}
        A_{\text{JCC},0}\xleft( y'_2 \right)
        \sqrt{ \frac{ S_{\text{JCC},0}\xleft( y'_1 - y_n \right)
               }
               { S_{\text{JCC},0}\xleft( y'_1 - y'_2 \right) 
                 \, S_{\text{JCC},0}\xleft( y_n - y'_2 \right)
               }
             }
\nonumber\\
   & = A_{\text{JCC},0}\xleft( -\infty \right)
        \sqrt{ \frac{ S_{\text{JCC},0}\xleft( +\infty - y_n \right)
               }
               { S_{\text{JCC},0}\xleft( +\infty - (-\infty) \right) 
                 \, S_{\text{JCC},0}\xleft( y_n - (-\infty) \right)
               }
             },
\end{align}
where the meaning of the second line is given by the limits in the
previous line.
The corresponding definition of $C_{\text{JCC},W}$ is
\begin{align}
\label{eq:CW.JCC.def}
  C_{\text{JCC,~unren},W}\xleft( y_n-y_2 \right)
   & = \lim_{{y'_1\to+\infty \atop y'_2\to-\infty}}
        S_{\text{JCC},0}\xleft( y'_1, y_2 \right)
        \sqrt{ \frac{ S_{\text{JCC},0}\xleft( y_n - y'_2 \right)
               }
               { S_{\text{JCC},0}\xleft( y'_1 - y'_2 \right) 
                 \, S_{\text{JCC},0}\xleft( y'_1 - y_n \right)
               }
             }
\nonumber\\
   & = S_{\text{JCC},0}\xleft( +\infty - y_2 \right)
        \sqrt{ \frac{ S_{\text{JCC},0}\xleft( y_n - (-\infty) \right)
               }
               { S_{\text{JCC},0}\xleft( +\infty - (-\infty) \right) 
                 \, S_{\text{JCC},0}\xleft( +\infty - y_n \right)
               }
             },
\end{align}
where the role of $y'_1$ and $y'_2$ is exchanged compared with the
definition of $A_{\text{JCC}}$.

Invariance under boosts in the $z$ direction
shows that the dependence of
$A_{\text{JCC}}$ on $k_A^+$ and $y_n$ is only via the combination
$(k_A^+)^2e^{-2y_n}$.  
Similarly, the dependence of $C_{\text{JCC},W}$
on $y_n$ and $y_2$ is only via $y_n-y_2$.  The derivation of
\eqref{eq:A0.JCC.fact} is simply a version of the proofs of
factorization for DIS and Drell-Yan cross sections that is given in
Ref.~\cite{Collins:2011qcdbook}.

We now apply the CS equation to 
obtain the $y_n$-dependence of $A_{\text{JCC}}\xleft(
2(k_A^+)^2e^{-2y_n} \right)$ and of 
$C_{\text{JCC},W}\xleft( y_n-y_2 \right)$ at fixed values of their
arguments.  The CS equations are
\begin{align}
\label{eq:csequation.A}
   \frac{\partial A_{\text{JCC,~unren}}\xleft( 2(k_A^+)^2e^{-2y_n} \right) }
        {\partial y_n}
   & = - A_{\text{JCC,~unren}}\xleft( 2(k_A^+)^2e^{-2y_n} \right)  \tilde{K}_{\text{unren}},
\\
\label{eq:csequation.C}
   \frac{\partial C_{\text{JCC,~unren},W}\xleft( y_n-y_2 \right) }
        {\partial y_n}
   & = C_{\text{JCC,~unren},W}\xleft( y_n-y_2 \right) \tilde{K}_{\text{unren}},
\end{align}
where $\tilde{K}$ is the CS kernel which depends on $b_T$ and $\mu$, but not the
other variables.  These equations have the immediate implication that
the product of $A$ and $C$ in \eqref{eq:A0.JCC.fact} is independent of
$y_n$.  The tilde in $\tilde{K}$ merely indicates that this is the
object in transverse position space, not in transverse momentum
space. 

Let $M$ be some chosen fixed 
reference mass scale.  
We use \eqref{eq:csequation.A} to express
$A_{\text{JCC}}\xleft( 2(k_A^+)^2e^{-2y_n} \right)$ in terms of its
value with $2(k_A^+)^2e^{-2y_n}$ replaced by 
$M^2$:
\begin{equation}
\label{eq:evolvedA}
   A_{\text{JCC,~unren}}\xleft( 2(k_A^+)^2e^{-2y_n} \right)
   = A_{\text{JCC,~unren}}\xleft( M^2 \right) 
      \exp \left[ \left( - \ln \frac{\sqrt{2} k_A^+}{M} - y_n \right)
        \tilde{K}_{\text{unren}} \right].
\end{equation}
Similarly,
\begin{equation}
\label{eq:evolvedC}
   C_{\text{JCC,~unren},W}\xleft( y_n-y_2 \right)
   = C_{\text{JCC,~unren},W}\xleft( 0 \right) 
      \exp \left[ \left( y_n-y_2 \right) \tilde{K}_{\text{unren}} \right].
\end{equation}
Applying these results to Eq.~\eqref{eq:A0.JCC.fact} allows us to
write $A_{\text{JCC},0}\xleft( y_2 \right)$ in terms of the collinear
factors at fixed values of their arguments, and thus to determine the
$y_2$ dependence of $A_{\text{JCC},0}\xleft( y_2 \right)$ for large
negative $y_2$:
\begin{equation}
\label{eq:A0.JCC.fact.1}
  A_{\text{JCC},0}\xleft( y_2 \right)
  = A_{\text{JCC,~unren}}\xleft( M^2 \right)
    \,
    \exp\xleft[ \left( \ln\frac{\sqrt{2}k_A^+}{M} -y_2 \right)
                \tilde{K}_{\text{unren}}
        \right]
    \, C_{\text{JCC,~unren},W}(0) + \mbox{error},
\end{equation}
with the error vanishing when $y_2\to-\infty$.

Exactly analogous steps show that the other unsubtracted quantities obey
\begin{equation}
\label{eq:B0.JCC.fact.1}
  B_{\text{JCC},0}\xleft( y_1 \right)
  = B_{\text{JCC,~unren}}\xleft( M^2 \right)
    \,
    \exp\xleft[ \left( y_1 - \ln\frac{\sqrt{2}k_B^-}{M} \right)
                \tilde{K}_{\text{unren}}
        \right]
    \, C_{\text{JCC,~unren},W}(0) + \mbox{error},
\end{equation}
and
\begin{equation}
\label{eq:S0.JCC.fact.1}
  S_{\text{JCC},0}\xleft( y_1,y_2 \right)
  = C_{\text{JCC,~unren},W}(0)
    \,
    \exp\xleft[ (y_1 - y_2) \tilde{K}_{\text{unren}} \right]
    \, C_{\text{JCC},W}(0) + \mbox{error}.
\end{equation}
The collinear factors associated with the Wilson lines are the same
for all the Wilson lines.  This is shown simply by exchanging the $+$
and $-$ coordinates.  

Inserting all of these results into $A_0B_0/S_0$ shows that the
dependence on $y_1$ and $y_2$ cancels, as do the factors of
$C_{\text{JCC},W}$.  
Therefore, the limit $y_1\to+\infty$ and
$y_2\to-\infty$ exists.  
It is just
\begin{align}
\label{eq:A0.B0.A0.JCC}
  \lim_{{y_1\to+\infty \atop y_2\to-\infty}}
        \frac{ A_{\text{JCC},0}\xleft( y_2 \right)
               B_{\text{JCC},0}\xleft( y_1 \right) }
             { S_{\text{JCC},0}\xleft( y_1,y_2 \right) }
  &= A_{\text{JCC,~unren}}\xleft( M^2 \right) \, B_{\text{JCC,~unren}}\xleft( M^2 \right) 
    \, \exp\xleft[ \ln \frac{Q^2}{M^2} \tilde{K}_{\text{unren}}  \right]
\nonumber\\
  & = A_{\text{unren}}\xleft( \zeta_A \right) \, B_{\text{unren}}\xleft( \zeta_B \right), 
\end{align}
given that $2k_A^+k_B^- = Q^2$ and $\zeta_A\zeta_B=Q^4$.  The last
result is exactly as in the CSS factorization formula, with the JCC
definition of the collinear factors.

\subsection{\texorpdfstring{$A_0$}{A0}, \texorpdfstring{$B_0$}{B0} and
  \texorpdfstring{$S_0$}{S0} in the EIS method}

For the EIS method, we again apply the JCC factorization again to
$A_0$, but with a changed regulator for the Wilson line.  Instead of
the non-light-like Wilson line of rapidity $y_2$ that is used in
\eqref{eq:A0.JCC.fact}, EIS use a $\delta$-regulator.  The steps for
deriving factorization equations like \eqref{eq:A0.JCC.fact} depend on
the wide separation of rapidity between the hadron of momentum $p_A$
and the Wilson line in $A_0$.  Thus the structure of factorization is
insensitive to the method for regulating the Wilson line 
(so long as it is a method consistent with factorization), and the
proof continues to apply.  The factorization formula corresponding to
\eqref{eq:A0.JCC.fact.1} is then
\begin{equation}
\label{eq:A0.EIS.fact}
  A_{\text{EIS},0}\xleft( \delta^+ \right)
  = A_{\text{JCC,~unren}}\xleft( 2(k_A^+)^2e^{-2y_n} \right)
    \, C_{\text{EIS,~unren},W}\xleft( y_n - \ln\delta^+ \right)
    + \mbox{error} ,
\end{equation}
where now the error vanishes as $\delta^+\to0$.  In obtaining this equation,
we have again used dimensional regularization.  This regulates any
soft and collinear divergences associated with massless quark and
gluon lines, and therefore we have removed the $\Delta$-regulator that EIS
use.  The collinear factor $A_{\text{JCC}}$ is the same as in
\eqref{eq:A0.JCC.fact}, because it is associated with the same
collinear subgraphs associated with the hadron.  But the different
regulator for the Wilson line implies that the $C_W$ factor is
different.

This factor, $C_{\text{EIS,~unren},W}\xleft( y_n-\ln\delta^+ \right)$, depends only
on the combination $y_n-\ln\delta^+$ but not on $y_n$ and $\delta^+$ separately,
because of boost invariance.  To see this result explicitly, we
examine the relevant Wilson line propagators:
\begin{equation}
   \frac{i}{-e^{-2 y_n } k^+ +  k^- + i 0}, 
\qquad
    \frac{i}{-k^+  + i \delta^+} .
\label{eq:twopropEIS}
\end{equation}
Suppose we simultaneously change $y_n$ to $y_n+\delta y$ and $\delta^+$ to
$\delta^+e^{\delta y}$, 
so that $y_n-\ln\delta^+$ is unchanged.  Then we change
variables on all loop momentum integrals in $C_{\text{EIS,~unren},W}$ by
corresponding boost factors: $k^+$ to $k^+e^{\delta y}$, and $k^-$ to
$k^-e^{-\delta y}$. This has effectively implemented a boost, and $C$ is
boost invariant, so its value is unchanged.

In accordance with the JCC factorization method, $C_{\text{EIS,~unren},W}$ is
defined by
\begin{align}
\label{eq:CW.EIS.def}
  C_{\text{EIS,~unren},W}\xleft( y_n-\ln\delta^+ \right)
   & = \lim_{{y'_1\to+\infty \atop y'_2\to-\infty}}
        C_{\text{EIS},0}\xleft( y'_1 - \ln\delta^+ \right)
        \sqrt{ \frac{ S_{\text{JCC},0}\xleft( y_n, y'_2 \right)
               }
               { S_{\text{JCC},0}\xleft( y'_1, y'_2 \right) 
                 \, S_{\text{JCC},0}\xleft( y'_1, y_n \right)
               }
             }
\nonumber\\
   & = C_{\text{EIS},0}\xleft( +\infty-\ln\delta^+ \right)
        \sqrt{ \frac{ S_{\text{JCC},0}\xleft( y_n, -\infty \right)
               }
               { S_{\text{JCC},0}\xleft( +\infty, -\infty \right) 
                 \, S_{\text{JCC},0}\xleft( +\infty, y_n \right)
               }
             }.
\end{align}
The bare factor is defined with one Wilson line regulated by the EIS
method and one regulated by the JCC method.

\begin{figure*}
  \centering
  \begin{tabular}{c@{\hspace*{10mm}}c}
       \VC{\includegraphics[scale=.4]{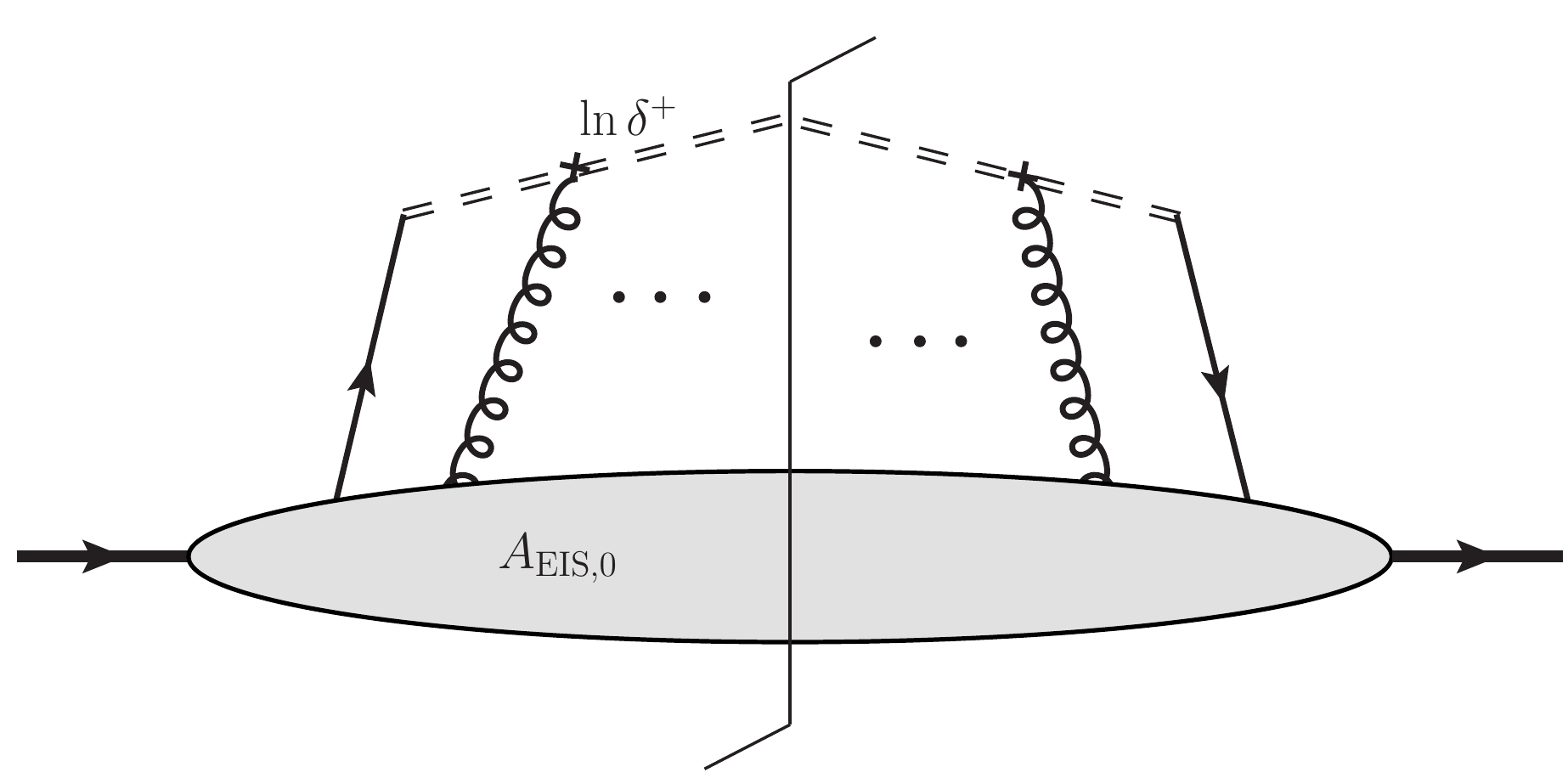}}
   & 
       \VC{\includegraphics[scale=.4]{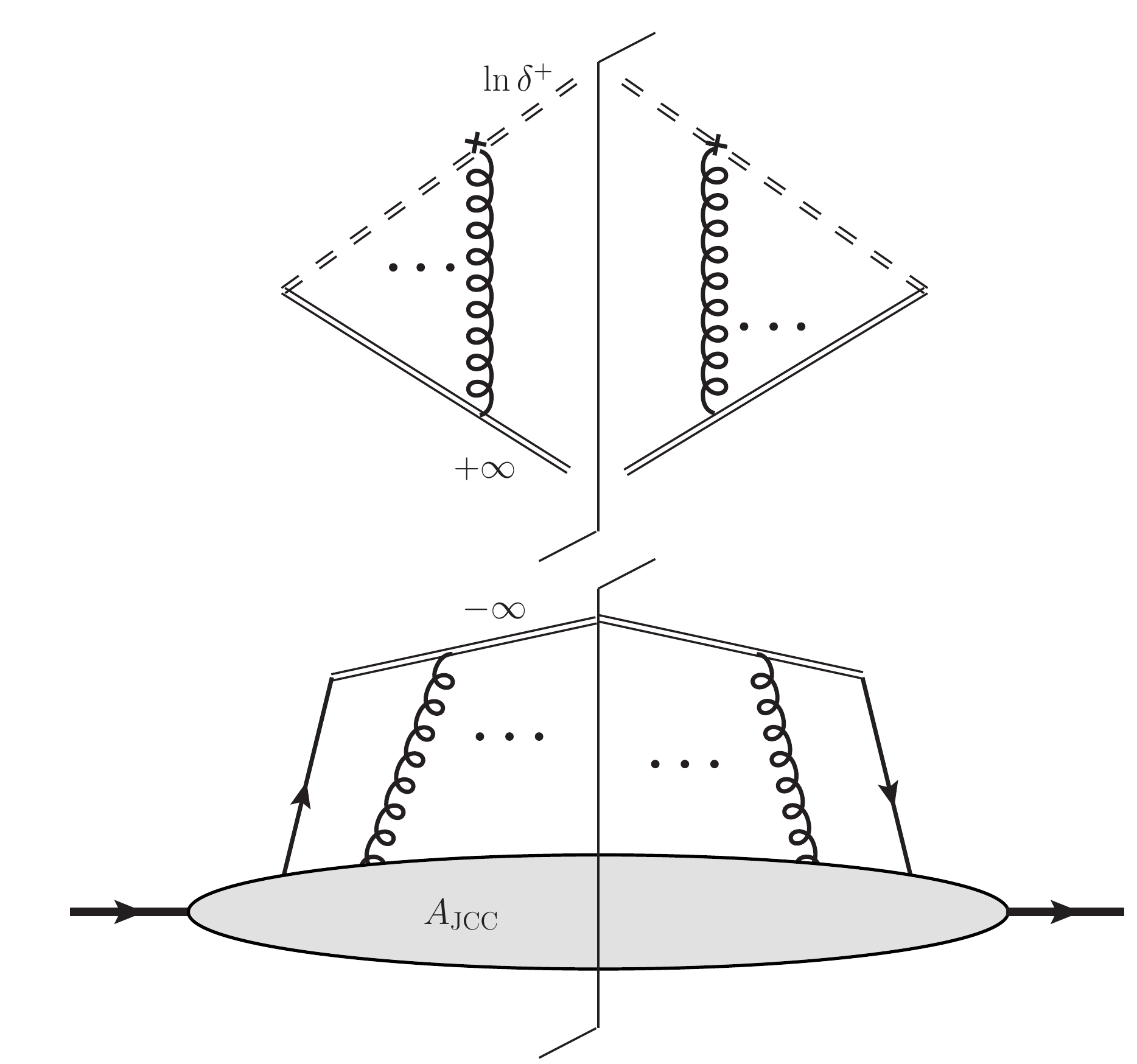}}
  \\
     (a) & (b)
  \end{tabular}
  \caption{For the factorization in \eqref{eq:A0.EIS.fact}:
     (a) the left-hand-side, and (b) the \emph{bare} collinear factors
     for the right-hand-side.  
     The dashed-double lines with cross-vertices represent the 
     $\delta$-regulated Wilson lines of 
     the EIS method.
  }
  \label{fig:A0EIS.fact}
\end{figure*}

The CS equations are the same as before, since they arise from the
same $S_{\text{JCC},0}$ factors under the square root.  Hence, instead
of \eqref{eq:A0.JCC.fact.1}, we get
\begin{equation}
\label{eq:A0.EIS.fact.1}
 A_{\text{EIS},0}\xleft( \delta^+ \right)
 = A_{\text{JCC,~unren}}\xleft( M^2 \right)
    \,
    \exp\xleft[ \left( \ln \frac{\sqrt{2}k_A^+}{M} 
                      - \ln \frac{ \delta^+ }{ M } \right)
                \tilde{K}_{\text{unren}}
        \right]
    \, C_{\text{EIS,~unren},W}(M) + \mbox{error}.
\end{equation}
This and the corresponding factorizations for $B_0$ and $S_0$ have
errors that vanish as $\delta^{\pm} \to 0$.  The collinear factor for
the Wilson line changed from that in \eqref{eq:A0.JCC.fact} and
\eqref{eq:A0.JCC.fact.1} because of the changed regulator.  
Note that the reference scale for the 
$C_{\text{EIS,~unren},W}(M)$ in Eq.~\eqref{eq:A0.EIS.fact.1} 
is now a mass $M$, since the 
regulator is a mass scale $\delta^+$.
But the
$A$ factor is the same as in the JCC method.

The formulas for $B_0$ and $S_0$ are
\begin{align}
\label{eq:B0.EIS.fact}
  B_{\text{EIS},0}\xleft( \delta^- \right)
  &= C_{\text{EIS,~unren},W}\xleft( -\ln\delta^- - y_n \right)
    \, B_{\text{JCC,~unren}}\xleft( 2(k_B^-)^2e^{2y_n} \right)
    + \mbox{error}
\nonumber\\
  &= 
    C_{\text{EIS,~unren},W}(M)
    \exp\xleft[ \left( \ln \frac{\sqrt{2}k_B^-}{M} 
                       - \ln \frac{ \delta^- }{ M } 
                \right)
                \tilde{K}_{\text{unren}}
        \right]
    \, B_{\text{JCC,~unren}}\xleft( M^2 \right) + \mbox{error},
\\
\label{eq:S0.EIS.fact}
  S_{\text{EIS},0}\xleft( \delta^-, \delta^+ \right)
  &= C_{\text{EIS,~unren},W}\xleft( -\ln\delta^- - y_n \right)
    \, C_{\text{EIS,~unren},W}\xleft( y_n -\ln\delta^+ \right)
    + \mbox{error}
\nonumber\\
  &= C_{\text{EIS,~unren},W}(M)
    \exp\xleft[ \left( \ln \frac{ M^2 }{ \delta^- \delta^+ }
                     \right)
                \tilde{K}_{\text{unren}}
        \right]
    \, C_{\text{EIS,~unren},W}(M) + \mbox{error},
\end{align}

Hence, the limit of $A_0B_0/S_0$ is the same as in the JCC
method: 
\begin{equation}
\label{eq:A0.B0.A0.EIS}
  \lim_{\delta^+, \delta^- \to 0}
   \frac{ A_{\text{EIS},0}\xleft( \delta^+ \right)
          B_{\text{EIS},0}\xleft( \delta^- \right)
        }
        { S_{\text{EIS},0}\xleft( \delta^-,\delta^+ \right) }
=
   A_{\text{JCC,~unren}}\xleft( M^2 \right) \, B_{\text{JCC,~unren}}\xleft( M^2 \right)
   \, \exp\xleft[ 
              \left( 
                  \ln \frac{ Q^2 }{ M^2 }
              \right)
              \tilde{K}_{\text{unren}}
          \right].
\end{equation}
As before, we have a cancellation of  the collinear factors for the
Wilson lines and of the dependence on the regulator parameters. Errors
go to zero in the limit that the regulator is removed.  Hence, we get
the same finite result with the EIS method as with the JCC method.

The above derivation used the JCC organization of factorization to
express the combination $A_0B_0/S_0$ of bare quantities in the EIS
method for the Drell-Yan cross section in terms of $A$ and $B$ in the
JCC scheme.  But this same product is the product of parton densities
in the EIS scheme.  Hence from Eq.~\eqref{eq:A0.B0.A0.EIS}, it follows
that the product $AB$ is the same in both methods.  The only question
now is whether the individual factors $A$ and $B$ are the same.

The only caveat in the above derivation arises because the derivation
of factorization uses gauge invariance heavily, at least beyond simple
one-loop graphs.  But 
when Wilson lines in gauge-invariant operators are replaced by
regulated Wilson lines in the EIS method, the resulting operators are
not gauge invariant.  This could potentially cause some problems.  It
needs to be investigated whether any problems actually occur.  It is
likely that any resulting corrections are suppressed by a power of the
regulator parameters $\delta^\pm$, and will not affect the final
results.

\section{The JCC Method and the Formula
  \texorpdfstring{$A=A_0/\sqrt{S_0}$}{A=A0/sqrt(S0)}} 
  \label{sec:JCCformula}

The unrenormalized collinear factors in the JCC method are defined by
\begin{align}
\label{eq:A.JCC.def}
  A_{\text{JCC,~unren}}\xleft( \zeta_A \right)
  & = \lim_{{y_1\to+\infty \atop y_2\to-\infty}}
        A_{\text{JCC},0}\xleft( y_2 \right)
        \sqrt{ \frac{ S_{\text{JCC},0}\xleft( y_1, y_n \right)
               }
               { S_{\text{JCC},0}\xleft( y_1, y_2 \right) 
                 \, S_{\text{JCC},0}\xleft( y_n, y_2 \right)
               }
             }
\\
\label{eq:B.JCC.def}
  B_{\text{JCC,~unren}}\xleft( \zeta_B \right)
  & = \lim_{{y_1\to+\infty \atop y_2\to-\infty}}
        B_{\text{JCC},0}\xleft( y_1 \right)
        \sqrt{ \frac{ S_{\text{JCC},0}\xleft( y_n, y_2 \right)
                }
                { S_{\text{JCC},0}\xleft( y_1, y_2 \right) 
                  \, S_{\text{JCC},0}\xleft( y_1, y_n \right)
                }
              },
\end{align}
where $\zeta_A$ and $\zeta_B$ are defined by Eqs.\ \eqref{eq:zeta.A} and
\eqref{eq:zeta.B}. Evidently the product of $A_{\text{JCC,~unren}}$
and $B_{\text{JCC,~unren}}$ is exactly (the appropriate limit of)
$A_0B_0/S_0$, a result we have already seen in Eqs.\
\eqref{eq:A0.B0.A0.JCC} and \eqref{eq:A0.B0.A0.EIS}.

In these formulas, the limits $y_1\to+\infty$, $y_2\to-\infty$ can be
taken independently.  But the rate at which the rapidities go to
infinity can also be coordinated.  Let us do this so that the rapidity
intervals are the same in both of the $S$ factors that have a $y_n$
argument, i.e., let us set
\begin{equation}
  y_1 = y_n + \Delta y,
\qquad
  y_2 = y_n - \Delta y.
\end{equation}
Then we take the limit $\Delta y \to \infty$.  
Now, 
\begin{equation}
  S_{\text{JCC},0}\xleft( y_1, y_n \right)
  = S_{\text{JCC},0}\xleft( y_n+\Delta y, y_n \right)
  = S_{\text{JCC},0}\xleft( y_n, y_n-\Delta y \right)
  = S_{\text{JCC},0}\xleft( y_n, y_2 \right),
\end{equation}
where the middle equality is obtained by exchanging the $+$ and $-$
coordinates, by applying charge conjugation to 
exchange the color charges
of the Wilson lines, and by then applying a boost.
After that, we apply UV renormalization.

We therefore obtain
\begin{align}
\label{eq:A.JCC.def.2}
  A_{\text{JCC}}\xleft( \zeta_A \right)
  & = \lim_{{y_1\to+\infty \atop y_2\to-\infty}}
        \frac{ A_{\text{JCC},0}\xleft( y_2 \right) }
             { \sqrt{ S_{\text{JCC},0}\xleft( y_1, y_2 \right) } }
      \times \mbox{UV renormalization factor},
\\
\label{eq:B.JCC.def.2}
  B_{\text{JCC}}\xleft( \zeta_B \right)
  & = \lim_{{y_1\to+\infty \atop y_2\to-\infty}}
        \frac{ B_{\text{JCC},0}\xleft( y_1 \right) }
             { \sqrt{ S_{\text{JCC},0}\xleft( y_1, y_2 \right) } }
      \times \mbox{UV renormalization factor},
\end{align}
where we require the limits to be taken with $\frac12(y_1+y_2) = y_n$
fixed.  
We can also write these with the limit of a single variable:
\begin{align}
\label{eq:A.JCC.def.3}
  A_{\text{JCC}}\xleft( \zeta_A \right)
  & = \lim_{{y_2\to-\infty}}
        \frac{ A_{\text{JCC},0}\xleft( y_2 \right) }
             { \sqrt{ S_{\text{JCC},0}\xleft( 2 y_n - y_2, y_2 \right) } }
      \times \mbox{UV renormalization factor},
\\
\label{eq:B.JCC.def.3}
  B_{\text{JCC}}\xleft( \zeta_B \right)
  & = \lim_{{y_1\to+\infty}}
        \frac{ B_{\text{JCC},0}\xleft( y_1 \right) }
             { \sqrt{ S_{\text{JCC},0}\xleft( y_1, 2 y_n - y_1\right) } }
      \times \mbox{UV renormalization factor}.
\end{align}
These formulas are simpler than the original JCC formulas, because they have two factors
instead of four, and they are of the form of the EIS definitions
except for the method of regulation of the light-like Wilson lines.

\section{EIS TMD PDFs have a \texorpdfstring{$\zeta$}{zeta} Parameter and are
  Equal to the JCC PDFs}
  \label{sec:EISzeta}

The EIS PDFs are defined by 
\begin{align}
\label{eq:A.EIS.def}
  A_{\text{EIS}}
  & = \lim_{\delta^+, \delta^- \to 0}
        \frac{ A_{\text{EIS},0}\xleft( \delta^+ \right) }
             { \sqrt{ S_{\text{EIS},0}\xleft( \delta^-, \delta^+ \right) } }
      \times \mbox{UV renormalization factor},
\\
\label{eq:B.EIS.def}
  B_{\text{EIS}}
  & = \lim_{\delta^+, \delta^- \to 0}
        \frac{ B_{\text{EIS},0}\xleft( \delta^- \right) }
             { \sqrt{ S_{\text{EIS},0}\xleft( \delta^-, \delta^+ \right) } }
      \times \mbox{UV renormalization factor}.
\end{align}
Compared with EIS's actual definition \cite[Eq.\
(2.13)]{GarciaEchevarria:2011rb}, we have changed the names of
factors, we have made explicit the regulator parameters, and we have
omitted the Fourier transform needed to get the TMD PDFs in transverse
momentum space.

In their statement of the definitions, EIS did not 
originally 
specify the relative rates
at which $\delta^+$ and $\delta^-$ go to zero. 
However, in their actual
calculations --- see \cite[Sec.\ 3.1]{GarciaEchevarria:2011rb}, they
set $\delta^+ = \delta^-$.  In fact, this is  not a boost
invariant condition, and, as we now show, it suffices to keep the ratio
$\delta^+/\delta^-$ fixed as $\delta^+$ and $\delta^-$ go to zero.  
From the perspective of factorization alone, 
any choice of fixed $\delta^+/\delta^-$ gives a distinct but 
equally valid definition in the limit of $\delta^+,\delta^- \to 0$.
The choice of $\delta^+ = \delta^-$, plus the choice to work in a frame where 
$k_A^+ = k_B^- = Q$, is what allows EIS to recover logarithms of $Q^2$ in their 
explicit calculation (e.g. \cite[Eq.\ 3.10]{GarciaEchevarria:2011rb}), despite the 
apparent absence of the scale $Q^2$ inside their TMD PDF definitions.

We apply \eqref{eq:A0.EIS.fact.1} and \eqref{eq:S0.EIS.fact} to
\eqref{eq:A.EIS.def}, and apply UV renormalization, to get
\begin{equation}
\label{eq:A.EIS.def.2}
  A_{\text{EIS}}
  = \lim_{\delta^+, \delta^- \to 0}
    A_{\text{JCC}}\xleft( M^2 \right)
    \,
    \exp\xleft[ \left( \ln \frac{\sqrt{2}k_A^+}{M} 
                      + \frac12 \ln \frac{ \delta^- }{ \delta^+ } \right)
                \tilde{K}
        \right].
\end{equation}
Evidently, the ratio $\delta^-/\delta^+$ finite and nonzero for the limit
to be well-defined.  
Equation~\eqref{eq:A.EIS.def.2} is \emph{exactly} equal to the 
JCC PDF, $A_{\text{JCC}}$, with
\begin{equation}
  \frac{ \delta^+ }{ \delta^- } = e^{2y_n}.
\end{equation}
Then, in exact correspondence to the JCC scheme, we define
\begin{equation}
\label{eq:zetaA}
  \zeta_A = 2(k_A^+)^2 e^{-2y_n} = 2 (k_A^+)^2 \frac{\delta^-}{\delta^+},
\end{equation}
and $A_{\text{EIS}}$ depends on $\zeta_A$.   Thus, the 
$\delta^+,\delta^-$ regulation scheme of 
Ref.~\cite{GarciaEchevarria:2011rb} is exactly equivalent to the JCC
scheme if we identify the rapidity parameter as
$y_n = \ln \sqrt{\delta^+/\delta^-}$.

An exactly corresponding equation applies to the other PDF:
\begin{align}
\label{eq:B.EIS.def.2}
  B_{\text{EIS}}\xleft( \zeta_B \right)
  & = \lim_{\delta^+, \delta^- \to 0}
    B_{\text{JCC}}\xleft( M^2 \right)
    \,
    \exp\xleft[ \left( \ln \frac{\sqrt{2}k_B^-}{M} 
                      + \frac12 \ln \frac{ \delta^+ }{ \delta^- } \right)
                \tilde{K}
        \right]
\nonumber\\
  & = 
    B_{\text{JCC}}\xleft( M^2 \right)
    \,
    \exp\xleft[ \left( \ln \frac{\sqrt{2}k_B^-}{M} + y_n \right)
                \tilde{K}
        \right],
\end{align}
with just an exchange of the roles of the $+$ and $-$ coordinates.
Now we have
\begin{equation}
\label{eq:zetaB}
  \zeta_B = 2(k_B^-)^2 e^{2y_n} = 2 (k_B^-)^2 \frac{\delta^+}{\delta^-}.
\end{equation}
\end{widetext}

The limit taken by EIS with $\delta^+ = \delta^-$ simply corresponds
to the case $y_n=0$, for which $\zeta_A=\zeta_B=Q^2$, provided that one
uses a frame where $k_A^+ = k_B^-$.  
Since the EIS PDFs have a $\zeta$ argument, and since they equal the
JCC PDFs, they must also obey a CS equation. 
In applications, the auxiliary parameters are to be evolved to values
that give good or optimal accuracy for the perturbative calculations
that are relevant for a given process at a given hard scale.
For the hard scattering, this entails setting $\mu$ and $\sqrt{\zeta}$ to
$Q$ (or to within a a finite factor of $Q$).  In this sense the two
auxiliary parameters can be treated as tied together.
(In fact only the
product $\zeta_A\zeta_B$ of the two $\zeta$s is relevant.)

Although EIS \cite{GarciaEchevarria:2011rb} claim not to use the CS
equation --- see e.g., their Sec.~10 --- they use an equivalent
result, obtained in their Sec.~5 for the $Q$-dependence of their TMD
pdfs.  They label this result resummation, but the result is exactly
equivalent to the use of the CS equation, the RG equation that is
already part of their formalism, and the small-$b$ expansion of TMD
pdfs.  The results are the same as in the CS method (in the JCC
version), as can be seen, for example, by comparing their Eq.~(5.7)
with Sec.~13.13.1 of Ref.~\cite{Collins:2011qcdbook}.  EIS's restrict
their results to the region $\Lambda \ll q_T \ll Q$.  However, with the
exception of the small-$b$ expansion, the results also apply for all
small $q_T$ as well, as is proved in Ref.~\cite{Collins:2011qcdbook}.

Furthermore, EIS 
use here that the logarithm of the TMD PDFs are linear in $\ln (Q^2 / \mu^2)$.  
(See \cite[Eq.\ (5.4)]{GarciaEchevarria:2011rb} and the discussion that follows.)
In fact, this linearity is a trivial consequence of the CS equation.

With the identifications in Eqs.~(\ref{eq:A.EIS.def.2}-\ref{eq:zetaB}), EIS 
find a one-loop anomalous
dimension \cite[Eq.\ (3.21,3.22)]{GarciaEchevarria:2011rb} that is the
same as the JCC one.

EIS, in a recent preprint \cite{Echevarria:2012qe}, advocate setting
$y_n = 0$ ($\delta^+ = \delta^-$) from the outset.
While this is legitimate, we find it non-optimal for the following reasons.
First, fixing a particular value for $y_n$ amounts to fixing 
a choice of reference frame with respect to boosts along the beam axis.  In the 
case of $y_n = 0$, it is the rest frame of the Drell-Yan lepton pair.  
This choice treats the hadrons asymmetrically because, although it corresponds 
to the parton rest frame, $k_A^+ = k_B^- = Q$, it is not the rest frame of the 
incoming hadrons, so the fractional momenta $x_A$ and $x_B$ will in general 
be very different, so it does not correspond to a particularly 
natural frame with regard to the TMD pdfs.  More importantly, 
setting $y_n$ to zero does not eliminate 
the CS evolution, but rather shifts it entirely into the dependence on $k_A^+$ 
in Eq.~\eqref{eq:zetaA} and $k_B^-$ in Eq.~\eqref{eq:zetaB}.  In fact, CS evolution 
is boost-independent, which is a particular advantage of the JCC approach, 
so we view it as preferable to leave the frame unspecified in the implementation 
of evolution.

\section{Different versions of  \texorpdfstring{\MSbar{}}{MSbar} renormalization}
\label{sec:MSbar}

However a possible problem is that calculations of the hard part do not agree:
Compare \cite[Eq.\ (6)]{Aybat:2011vb} with \cite[Eq.\
(7.5)]{GarciaEchevarria:2011rb}, where the $\pi^2$ terms are different
by a factor of $7/6$.
We suspect that this is due to a difference in the definitions
of the \MSbar{} scheme for UV renormalization.  JCC
\cite[Sec.\ 3.3.2]{Collins:2011qcdbook}
defines one-loop
counterterms to have the structure
\begin{equation}
  \label{eq:MSbar.JCC}
  S_\epsilon \times \mbox{sum of poles in $\epsilon$},
\end{equation}
where
\begin{equation}
  \label{eq:S.epsilon.JCC}
  S_\epsilon = \frac{ (4\pi)^\epsilon }{ \Gamma(1-\epsilon) }.
\end{equation}
A more commonly used definition would replace $S_\epsilon$ by
$(4\pi e^{-\gamma_E})^\epsilon$, which differs by a term of order $\epsilon^2$.  Normal UV
renormalization only needs at most one factor of $1/\epsilon$ per loop, and
the change in the definition of the \MSbar{} scheme is irrelevant for
such quantities.  But this is no longer true when there are UV
divergences with a double pole $1/\epsilon^2$ per loop, as in TMD pdfs with
the EIS and JCC definitions.  The modified definition
Eq.~(\ref{eq:S.epsilon.JCC}) was made specifically with this situation
in mind, since the factor $S_\epsilon$ is common to all one-loop
transverse-momentum integrals in dimensional regularization.
The use of (\ref{eq:S.epsilon.JCC}) completes the simplifications of
the results of one-loop calculations provided by the \MSbar{} scheme.
We have checked that the difference in the coefficient of $\pi^2$ is
consistent with this change in the definition of the \MSbar{} scheme.
However, not enough details are given in
\cite{GarciaEchevarria:2011rb} for us to verify definitively that this
is the correct explanation of the difference between the reported
hard-scattering coefficients.

\section{Summary}
\label{sec:summary}

In this article, we have shown 
that two seemingly different methods of formulating 
TMD-factorization --- one originating in ideas from SCET, and one based on 
subtractive methods of pQCD --- are actually equivalent.  That the same 
conclusions are found from two seemingly different approaches  
provides important support for the general validity of the
TMD-factorization formalism.  

To clarify this equivalence, and to understand the origin of the 
superficial disparity between the two 
approaches, 
it is necessary to comment on some differences between what we have referred to   
as the EIS method in Eq.~\eqref{eq:AB.0.EIS} and the 
original presentation of the EIS formalism in Ref.~\cite{GarciaEchevarria:2011rb}:
We have naturally assumed it to be implicit in Ref.~\cite{GarciaEchevarria:2011rb} 
that the limit $\delta^+, \delta^- \to 0$ for the regulators of the Wilson lines is
to be taken, although this appears not to be explicitly stated.
Now the calculations in Ref.~\cite{GarciaEchevarria:2011rb} 
rely on separate regulators $\Delta^+$ and $\Delta^-$ for the quark lines, as
well as regulators $\delta^+$ and $\delta^-$ for the Wilson lines.
The two kinds of regulator were tied by the conditions $\Delta^+ = \delta^+
p_B^-$ and $\Delta^- = \delta^-p_A^+$, so that in massless
calculations, an appropriate soft approximation is 
valid.  The $\Delta$ regulators cut off infrared divergences associated with
the quark lines.  If some other cutoff is imposed, for example, by
nonzero masses, then the $\Delta$ regulator can be removed without being
tied to the $\delta$ regulator. 
In our general 
discussion, we have assumed that some other infrared cutoff is 
present throughout, and we have set $\Delta^+ = \Delta^- = 0$ from the
outset.
We note that the consistency of $\delta$-regulator formalism is derived 
in Ref.~\cite{GarciaEchevarria:2011rb} from an examination of 
one-loop graphs with on-shell external partons.  It is not clear to us
how complete the derivations Ref.~\cite{GarciaEchevarria:2011rb} are
beyond that order.
For the purposes of this paper,
we have taken 
as our starting point the assumption that overall factorization with the $\delta$-regulators remains consistent in 
a general derivation at arbitrary order.

We made one other explicit extension to the EIS definition.  EIS
\cite{GarciaEchevarria:2011rb} imposed the condition $\delta^+=\delta^-$.  But
this is boost dependent.   We have instead allowed the ratio $\delta^+/\delta^-$
to remain unspecified in Eq.~\eqref{eq:AB.0.EIS}  (its variation is responsible for
the CS equation), and have observed that the EIS definition continues
to work in this case\footnote{We thank Markus Diehl
  for bringing to our attention the issues discussed in this and the previoius paragraph.}.

Given the consistency assumption, we find not only that the methods (EIS and JCC) are 
equivalent, but that the different definitions give exactly equal TMD pdfs,
with the exception of different definitions of the \MSbar{}
scheme, as explained around Eq.~(\ref{eq:S.epsilon.JCC}).
A criticism that is sometimes made of the CSS formalism 
is that it is prohibitively difficult or 
complicated for practical use, particular in the context of probing non-perturbative structure.  
However, we believe that the most recent version, presented in Ref.~\cite{Collins:2011qcdbook},
has eliminated the main issues responsible for such difficulties, and now provides a 
maximally user-friendly TMD formalism.  
At least in the manual one-loop calculations we have performed, the
length of the calculations is little different.
The remaining complexities are those of 
ordinary Feynman graph calculations. 

We conclude that since approaches to TMD factorization that are
apparently widely different nevertheless give the same TMD pdfs, this
supports the idea that there is a particularly natural kind of
preferred 
definition of TMD pdfs in QCD.  This gives good current and near-future
prospects~\cite{Aidala:2012mv,Anselmino:2012aa,Anselmino:2012re,Guzzi:2012jc,Aybat:2011ta} for 
applications of TMD-factorization in a broad variety 
of phenomenological studies.  There remain differences in how the
different groups analyze the non-perturbative large-$b$ region in the
solution of the evolution equations.  We leave the analysis of these
differences to future work.  The TMD pdfs themselves, once given an
operator definition, are unambiguous entities in QCD.

However, for precise phenomenological work there needs to be agreement
on the definition of the \MSbar{} scheme; differences in its
definition matter for TMD pdfs even though they do not affect simpler
quantities (where the UV divergences give only at most one factor of
$1/\epsilon$ per loop).  But this issue is not a matter of principle.


\section*{Acknowledgments}
We are very grateful to 
M.~Diehl,
M.~G.~Echevarria, A.~Idilbi, A. Prokudin,
I.~Scimemi and A.~Sch\"afer for useful discussions.
Feynman diagrams were produced using Jaxodraw~\cite{Binosi:2003yf,Binosi:2008ig}.
J.~C.~Collins was supported by the U.S. Department of 
Energy under the grant number DE-SC0008745
and T.~C.~Rogers was supported by the National Science Foundation, grant PHY-0969739.

\bibliography{jcctcr}

\providecommand{\noopsort}[1]{}
\begin{thebibliography}{26}%
\makeatletter
\providecommand \@ifxundefined [1]{%
 \@ifx{#1\undefined}
}%
\providecommand \@ifnum [1]{%
 \ifnum #1\expandafter \@firstoftwo
 \else \expandafter \@secondoftwo
 \fi
}%
\providecommand \@ifx [1]{%
 \ifx #1\expandafter \@firstoftwo
 \else \expandafter \@secondoftwo
 \fi
}%
\providecommand \natexlab [1]{#1}%
\providecommand \enquote  [1]{``#1''}%
\providecommand \bibnamefont  [1]{#1}%
\providecommand \bibfnamefont [1]{#1}%
\providecommand \citenamefont [1]{#1}%
\providecommand \href@noop [0]{\@secondoftwo}%
\providecommand \href [0]{\begingroup \@sanitize@url \@href}%
\providecommand \@href[1]{\@@startlink{#1}\@@href}%
\providecommand \@@href[1]{\endgroup#1\@@endlink}%
\providecommand \@sanitize@url [0]{\catcode `\\12\catcode `\$12\catcode
  `\&12\catcode `\#12\catcode `\^12\catcode `\_12\catcode `\%12\relax}%
\providecommand \@@startlink[1]{}%
\providecommand \@@endlink[0]{}%
\providecommand \url  [0]{\begingroup\@sanitize@url \@url }%
\providecommand \@url [1]{\endgroup\@href {#1}{\urlprefix }}%
\providecommand \urlprefix  [0]{URL }%
\providecommand \Eprint [0]{\href }%
\providecommand \doibase [0]{http://dx.doi.org/}%
\providecommand \selectlanguage [0]{\@gobble}%
\providecommand \bibinfo  [0]{\@secondoftwo}%
\providecommand \bibfield  [0]{\@secondoftwo}%
\providecommand \translation [1]{[#1]}%
\providecommand \BibitemOpen [0]{}%
\providecommand \bibitemStop [0]{}%
\providecommand \bibitemNoStop [0]{.\EOS\space}%
\providecommand \EOS [0]{\spacefactor3000\relax}%
\providecommand \BibitemShut  [1]{\csname bibitem#1\endcsname}%
\let\auto@bib@innerbib\@empty
\bibitem [{\citenamefont {Collins}\ and\ \citenamefont
  {Soper}(1982)}]{Collins:1981uw}%
  \BibitemOpen
  \bibfield  {author} {\bibinfo {author} {\bibfnamefont {J.~C.}\ \bibnamefont
  {Collins}}\ and\ \bibinfo {author} {\bibfnamefont {D.~E.}\ \bibnamefont
  {Soper}},\ }\href@noop {} {\bibfield  {journal} {\bibinfo  {journal} {Nucl.
  Phys.}\ }\textbf {\bibinfo {volume} {B194}},\ \bibinfo {pages} {445}
  (\bibinfo {year} {1982})}\BibitemShut {NoStop}%
\bibitem [{\citenamefont {Collins}\ \emph {et~al.}(1985)\citenamefont
  {Collins}, \citenamefont {Soper},\ and\ \citenamefont
  {Sterman}}]{Collins:1984kg}%
  \BibitemOpen
  \bibfield  {author} {\bibinfo {author} {\bibfnamefont {J.~C.}\ \bibnamefont
  {Collins}}, \bibinfo {author} {\bibfnamefont {D.~E.}\ \bibnamefont {Soper}},
  \ and\ \bibinfo {author} {\bibfnamefont {G.}~\bibnamefont {Sterman}},\
  }\href@noop {} {\bibfield  {journal} {\bibinfo  {journal} {Nucl. Phys.}\
  }\textbf {\bibinfo {volume} {B250}},\ \bibinfo {pages} {199} (\bibinfo {year}
  {1985})}\BibitemShut {NoStop}%
\bibitem [{\citenamefont {Garcia-Echevarria}\ \emph
  {et~al.}(2011{\natexlab{a}})\citenamefont {Garcia-Echevarria}, \citenamefont
  {Idilbi},\ and\ \citenamefont {Scimemi}}]{GarciaEchevarria:2011rb}%
  \BibitemOpen
  \bibfield  {author} {\bibinfo {author} {\bibfnamefont {M.}~\bibnamefont
  {Garcia-Echevarria}}, \bibinfo {author} {\bibfnamefont {A.}~\bibnamefont
  {Idilbi}}, \ and\ \bibinfo {author} {\bibfnamefont {I.}~\bibnamefont
  {Scimemi}},\ }\href@noop {} {\  (\bibinfo {year} {2011}{\natexlab{a}})},\
  \Eprint {http://arxiv.org/abs/1111.4996} {arXiv:1111.4996 [hep-ph]}
  \BibitemShut {NoStop}%
\bibitem [{\citenamefont {Echevarria}\ \emph
  {et~al.}(2012{\natexlab{a}})\citenamefont {Echevarria}, \citenamefont
  {Idilbi}, \citenamefont {Schafer},\ and\ \citenamefont
  {Scimemi}}]{Echevarria:2012pw}%
  \BibitemOpen
  \bibfield  {author} {\bibinfo {author} {\bibfnamefont {M.~G.}\ \bibnamefont
  {Echevarria}}, \bibinfo {author} {\bibfnamefont {A.}~\bibnamefont {Idilbi}},
  \bibinfo {author} {\bibfnamefont {A.}~\bibnamefont {Schafer}}, \ and\
  \bibinfo {author} {\bibfnamefont {I.}~\bibnamefont {Scimemi}},\ }\href@noop
  {} {\  (\bibinfo {year} {2012}{\natexlab{a}})},\ \Eprint
  {http://arxiv.org/abs/1208.1281} {arXiv:1208.1281 [hep-ph]} \BibitemShut
  {NoStop}%
\bibitem [{\citenamefont {Collins}(2011)}]{Collins:2011qcdbook}%
  \BibitemOpen
  \bibfield  {author} {\bibinfo {author} {\bibfnamefont {J.~C.}\ \bibnamefont
  {Collins}},\ }\href@noop {} {\emph {\bibinfo {title} {Foundations of
  Perturbative QCD}}}\ (\bibinfo  {publisher} {Cambridge University Press},\
  \bibinfo {address} {Cambridge},\ \bibinfo {year} {2011})\BibitemShut
  {NoStop}%
\bibitem [{\citenamefont {Korchemsky}\ and\ \citenamefont
  {Radyushkin}(1987)}]{Korchemsky:1987wg}%
  \BibitemOpen
  \bibfield  {author} {\bibinfo {author} {\bibfnamefont {G.}~\bibnamefont
  {Korchemsky}}\ and\ \bibinfo {author} {\bibfnamefont {A.}~\bibnamefont
  {Radyushkin}},\ }\href {\doibase 10.1016/0550-3213(87)90277-X} {\bibfield
  {journal} {\bibinfo  {journal} {Nucl. Phys.}\ }\textbf {\bibinfo {volume}
  {B283}},\ \bibinfo {pages} {342} (\bibinfo {year} {1987})}\BibitemShut
  {NoStop}%
\bibitem [{\citenamefont {Echevarria}\ \emph
  {et~al.}(2012{\natexlab{b}})\citenamefont {Echevarria}, \citenamefont
  {Idilbi},\ and\ \citenamefont {Scimemi}}]{Echevarria:2012qe}%
  \BibitemOpen
  \bibfield  {author} {\bibinfo {author} {\bibfnamefont {M.~G.}\ \bibnamefont
  {Echevarria}}, \bibinfo {author} {\bibfnamefont {A.}~\bibnamefont {Idilbi}},
  \ and\ \bibinfo {author} {\bibfnamefont {I.}~\bibnamefont {Scimemi}},\
  }\href@noop {} {\  (\bibinfo {year} {2012}{\natexlab{b}})},\ \Eprint
  {http://arxiv.org/abs/1209.3892} {arXiv:1209.3892 [hep-ph]} \BibitemShut
  {NoStop}%
\bibitem [{\citenamefont {Ji}\ \emph {et~al.}(2005)\citenamefont {Ji},
  \citenamefont {Ma},\ and\ \citenamefont {Yuan}}]{Ji:2004wu}%
  \BibitemOpen
  \bibfield  {author} {\bibinfo {author} {\bibfnamefont {X.-D.}\ \bibnamefont
  {Ji}}, \bibinfo {author} {\bibfnamefont {J.-P.}\ \bibnamefont {Ma}}, \ and\
  \bibinfo {author} {\bibfnamefont {F.}~\bibnamefont {Yuan}},\ }\href {\doibase
  10.1103/PhysRevD.71.034005} {\bibfield  {journal} {\bibinfo  {journal} {Phys.
  Rev.}\ }\textbf {\bibinfo {volume} {D71}},\ \bibinfo {pages} {034005}
  (\bibinfo {year} {2005})},\ \Eprint {http://arxiv.org/abs/hep-ph/0404183}
  {arXiv:hep-ph/0404183} \BibitemShut {NoStop}%
\bibitem [{\citenamefont {Ji}\ \emph {et~al.}(2004)\citenamefont {Ji},
  \citenamefont {Ma},\ and\ \citenamefont {Yuan}}]{Ji:2004xq}%
  \BibitemOpen
  \bibfield  {author} {\bibinfo {author} {\bibfnamefont {X.-D.}\ \bibnamefont
  {Ji}}, \bibinfo {author} {\bibfnamefont {J.-P.}\ \bibnamefont {Ma}}, \ and\
  \bibinfo {author} {\bibfnamefont {F.}~\bibnamefont {Yuan}},\ }\href {\doibase
  10.1016/j.physletb.2004.07.026} {\bibfield  {journal} {\bibinfo  {journal}
  {Phys. Lett.}\ }\textbf {\bibinfo {volume} {B597}},\ \bibinfo {pages} {299}
  (\bibinfo {year} {2004})},\ \Eprint {http://arxiv.org/abs/hep-ph/0405085}
  {arXiv:hep-ph/0405085 [hep-ph]} \BibitemShut {NoStop}%
\bibitem [{\citenamefont {Cherednikov}\ and\ \citenamefont
  {Stefanis}(2008)}]{Cherednikov:2007tw}%
  \BibitemOpen
  \bibfield  {author} {\bibinfo {author} {\bibfnamefont {I.}~\bibnamefont
  {Cherednikov}}\ and\ \bibinfo {author} {\bibfnamefont {N.}~\bibnamefont
  {Stefanis}},\ }\href {\doibase 10.1103/PhysRevD.77.094001} {\bibfield
  {journal} {\bibinfo  {journal} {Phys. Rev.}\ }\textbf {\bibinfo {volume}
  {D77}},\ \bibinfo {pages} {094001} (\bibinfo {year} {2008})},\ \Eprint
  {http://arxiv.org/abs/0710.1955} {arXiv:0710.1955 [hep-ph]} \BibitemShut
  {NoStop}%
\bibitem [{\citenamefont {Cherednikov}\ and\ \citenamefont
  {Stefanis}(2009)}]{Cherednikov:2009wk}%
  \BibitemOpen
  \bibfield  {author} {\bibinfo {author} {\bibfnamefont {I.~O.}\ \bibnamefont
  {Cherednikov}}\ and\ \bibinfo {author} {\bibfnamefont {N.~G.}\ \bibnamefont
  {Stefanis}},\ }\href {\doibase 10.1103/PhysRevD.80.054008} {\bibfield
  {journal} {\bibinfo  {journal} {Phys. Rev.}\ }\textbf {\bibinfo {volume}
  {D80}},\ \bibinfo {pages} {054008} (\bibinfo {year} {2009})},\ \Eprint
  {http://arxiv.org/abs/0904.2727} {arXiv:0904.2727 [hep-ph]} \BibitemShut
  {NoStop}%
\bibitem [{\citenamefont {Mantry}\ and\ \citenamefont
  {Petriello}(2011)}]{Mantry:2010bi}%
  \BibitemOpen
  \bibfield  {author} {\bibinfo {author} {\bibfnamefont {S.}~\bibnamefont
  {Mantry}}\ and\ \bibinfo {author} {\bibfnamefont {F.}~\bibnamefont
  {Petriello}},\ }\href {\doibase 10.1103/PhysRevD.84.014030} {\bibfield
  {journal} {\bibinfo  {journal} {Phys. Rev.}\ }\textbf {\bibinfo {volume}
  {D84}},\ \bibinfo {pages} {014030} (\bibinfo {year} {2011})},\ \Eprint
  {http://arxiv.org/abs/1011.0757} {arXiv:1011.0757 [hep-ph]} \BibitemShut
  {NoStop}%
\bibitem [{\citenamefont {Becher}\ and\ \citenamefont
  {Neubert}(2011)}]{Becher:2010tm}%
  \BibitemOpen
  \bibfield  {author} {\bibinfo {author} {\bibfnamefont {T.}~\bibnamefont
  {Becher}}\ and\ \bibinfo {author} {\bibfnamefont {M.}~\bibnamefont
  {Neubert}},\ }\href {\doibase 10.1140/epjc/s10052-011-1665-7} {\bibfield
  {journal} {\bibinfo  {journal} {Eur. Phys. J.}\ }\textbf {\bibinfo {volume}
  {C71}},\ \bibinfo {pages} {1665} (\bibinfo {year} {2011})},\ \Eprint
  {http://arxiv.org/abs/1007.4005} {arXiv:1007.4005 [hep-ph]} \BibitemShut
  {NoStop}%
\bibitem [{\citenamefont {Chay}\ and\ \citenamefont {Kim}(2012)}]{Chay:2012mh}%
  \BibitemOpen
  \bibfield  {author} {\bibinfo {author} {\bibfnamefont {J.}~\bibnamefont
  {Chay}}\ and\ \bibinfo {author} {\bibfnamefont {C.}~\bibnamefont {Kim}},\
  }\href@noop {} {\  (\bibinfo {year} {2012})},\ \Eprint
  {http://arxiv.org/abs/1208.0662} {arXiv:1208.0662 [hep-ph]} \BibitemShut
  {NoStop}%
\bibitem [{\citenamefont {Belitsky}\ \emph {et~al.}(2003)\citenamefont
  {Belitsky}, \citenamefont {Ji},\ and\ \citenamefont
  {Yuan}}]{Belitsky:2002sm}%
  \BibitemOpen
  \bibfield  {author} {\bibinfo {author} {\bibfnamefont {A.~V.}\ \bibnamefont
  {Belitsky}}, \bibinfo {author} {\bibfnamefont {X.}~\bibnamefont {Ji}}, \ and\
  \bibinfo {author} {\bibfnamefont {F.}~\bibnamefont {Yuan}},\ }\href@noop {}
  {\bibfield  {journal} {\bibinfo  {journal} {Nucl. Phys.}\ }\textbf {\bibinfo
  {volume} {B656}},\ \bibinfo {pages} {165} (\bibinfo {year} {2003})},\ \Eprint
  {http://arxiv.org/abs/hep-ph/0208038} {hep-ph/0208038} \BibitemShut {NoStop}%
\bibitem [{\citenamefont {Ji}\ and\ \citenamefont {Yuan}(2002)}]{Ji:2002aa}%
  \BibitemOpen
  \bibfield  {author} {\bibinfo {author} {\bibfnamefont {X.-D.}\ \bibnamefont
  {Ji}}\ and\ \bibinfo {author} {\bibfnamefont {F.}~\bibnamefont {Yuan}},\
  }\href@noop {} {\bibfield  {journal} {\bibinfo  {journal} {Phys. Lett.}\
  }\textbf {\bibinfo {volume} {B543}},\ \bibinfo {pages} {66} (\bibinfo {year}
  {2002})},\ \Eprint {http://arxiv.org/abs/hep-ph/0206057} {hep-ph/0206057}
  \BibitemShut {NoStop}%
\bibitem [{\citenamefont {Garcia-Echevarria}\ \emph
  {et~al.}(2011{\natexlab{b}})\citenamefont {Garcia-Echevarria}, \citenamefont
  {Idilbi},\ and\ \citenamefont {Scimemi}}]{GarciaEchevarria:2011md}%
  \BibitemOpen
  \bibfield  {author} {\bibinfo {author} {\bibfnamefont {M.}~\bibnamefont
  {Garcia-Echevarria}}, \bibinfo {author} {\bibfnamefont {A.}~\bibnamefont
  {Idilbi}}, \ and\ \bibinfo {author} {\bibfnamefont {I.}~\bibnamefont
  {Scimemi}},\ }\href {\doibase 10.1103/PhysRevD.84.011502} {\bibfield
  {journal} {\bibinfo  {journal} {Phys.Rev.}\ }\textbf {\bibinfo {volume}
  {D84}},\ \bibinfo {pages} {011502} (\bibinfo {year} {2011}{\natexlab{b}})},\
  \Eprint {http://arxiv.org/abs/1104.0686} {arXiv:1104.0686 [hep-ph]}
  \BibitemShut {NoStop}%
\bibitem [{\citenamefont {Idilbi}\ \emph {et~al.}(2004)\citenamefont {Idilbi},
  \citenamefont {Ji}, \citenamefont {Ma},\ and\ \citenamefont
  {Yuan}}]{Idilbi:2004vb}%
  \BibitemOpen
  \bibfield  {author} {\bibinfo {author} {\bibfnamefont {A.}~\bibnamefont
  {Idilbi}}, \bibinfo {author} {\bibfnamefont {X.-d.}\ \bibnamefont {Ji}},
  \bibinfo {author} {\bibfnamefont {J.-P.}\ \bibnamefont {Ma}}, \ and\ \bibinfo
  {author} {\bibfnamefont {F.}~\bibnamefont {Yuan}},\ }\href {\doibase
  10.1103/PhysRevD.70.074021} {\bibfield  {journal} {\bibinfo  {journal} {Phys.
  Rev.}\ }\textbf {\bibinfo {volume} {D70}},\ \bibinfo {pages} {074021}
  (\bibinfo {year} {2004})},\ \Eprint {http://arxiv.org/abs/hep-ph/0406302}
  {arXiv:hep-ph/0406302} \BibitemShut {NoStop}%
\bibitem [{\citenamefont {Aybat}\ and\ \citenamefont
  {Rogers}(2011)}]{Aybat:2011vb}%
  \BibitemOpen
  \bibfield  {author} {\bibinfo {author} {\bibfnamefont {S.}~\bibnamefont
  {Aybat}}\ and\ \bibinfo {author} {\bibfnamefont {T.}~\bibnamefont {Rogers}},\
  }\href@noop {} {\  (\bibinfo {year} {2011})},\ \Eprint
  {http://arxiv.org/abs/1107.3973} {arXiv:1107.3973 [hep-ph]} \BibitemShut
  {NoStop}%
\bibitem [{\citenamefont {Aidala}\ \emph {et~al.}(2012)\citenamefont {Aidala},
  \citenamefont {Bass}, \citenamefont {Hasch},\ and\ \citenamefont
  {Mallot}}]{Aidala:2012mv}%
  \BibitemOpen
  \bibfield  {author} {\bibinfo {author} {\bibfnamefont {C.~A.}\ \bibnamefont
  {Aidala}}, \bibinfo {author} {\bibfnamefont {S.~D.}\ \bibnamefont {Bass}},
  \bibinfo {author} {\bibfnamefont {D.}~\bibnamefont {Hasch}}, \ and\ \bibinfo
  {author} {\bibfnamefont {G.~K.}\ \bibnamefont {Mallot}},\ }\href@noop {} {\
  (\bibinfo {year} {2012})},\ \Eprint {http://arxiv.org/abs/1209.2803}
  {arXiv:1209.2803 [hep-ph]} \BibitemShut {NoStop}%
\bibitem [{\citenamefont {Anselmino}\ \emph
  {et~al.}(2012{\natexlab{a}})\citenamefont {Anselmino}, \citenamefont
  {Boglione},\ and\ \citenamefont {Melis}}]{Anselmino:2012aa}%
  \BibitemOpen
  \bibfield  {author} {\bibinfo {author} {\bibfnamefont {M.}~\bibnamefont
  {Anselmino}}, \bibinfo {author} {\bibfnamefont {M.}~\bibnamefont {Boglione}},
  \ and\ \bibinfo {author} {\bibfnamefont {S.}~\bibnamefont {Melis}},\ }\href
  {\doibase 10.1103/PhysRevD.86.014028} {\bibfield  {journal} {\bibinfo
  {journal} {Phys. Rev.}\ }\textbf {\bibinfo {volume} {D86}},\ \bibinfo {pages}
  {014028} (\bibinfo {year} {2012}{\natexlab{a}})},\ \Eprint
  {http://arxiv.org/abs/1204.1239} {arXiv:1204.1239 [hep-ph]} \BibitemShut
  {NoStop}%
\bibitem [{\citenamefont {Anselmino}\ \emph
  {et~al.}(2012{\natexlab{b}})\citenamefont {Anselmino}, \citenamefont
  {Boglione},\ and\ \citenamefont {Melis}}]{Anselmino:2012re}%
  \BibitemOpen
  \bibfield  {author} {\bibinfo {author} {\bibfnamefont {M.}~\bibnamefont
  {Anselmino}}, \bibinfo {author} {\bibfnamefont {M.}~\bibnamefont {Boglione}},
  \ and\ \bibinfo {author} {\bibfnamefont {S.}~\bibnamefont {Melis}},\
  }\href@noop {} {\  (\bibinfo {year} {2012}{\natexlab{b}})},\ \Eprint
  {http://arxiv.org/abs/1209.1541} {arXiv:1209.1541 [hep-ph]} \BibitemShut
  {NoStop}%
\bibitem [{\citenamefont {Guzzi}\ and\ \citenamefont
  {Nadolsky}(2012)}]{Guzzi:2012jc}%
  \BibitemOpen
  \bibfield  {author} {\bibinfo {author} {\bibfnamefont {M.}~\bibnamefont
  {Guzzi}}\ and\ \bibinfo {author} {\bibfnamefont {P.~M.}\ \bibnamefont
  {Nadolsky}},\ }\href@noop {} {\  (\bibinfo {year} {2012})},\ \Eprint
  {http://arxiv.org/abs/1209.1252} {arXiv:1209.1252 [hep-ph]} \BibitemShut
  {NoStop}%
\bibitem [{\citenamefont {Aybat}\ \emph {et~al.}(2012)\citenamefont {Aybat},
  \citenamefont {Prokudin},\ and\ \citenamefont {Rogers}}]{Aybat:2011ta}%
  \BibitemOpen
  \bibfield  {author} {\bibinfo {author} {\bibfnamefont {S.~M.}\ \bibnamefont
  {Aybat}}, \bibinfo {author} {\bibfnamefont {A.}~\bibnamefont {Prokudin}}, \
  and\ \bibinfo {author} {\bibfnamefont {T.~C.}\ \bibnamefont {Rogers}},\
  }\href {\doibase 10.1103/PhysRevLett.108.242003} {\bibfield  {journal}
  {\bibinfo  {journal} {Phys. Rev. Lett.}\ }\textbf {\bibinfo {volume} {108}},\
  \bibinfo {pages} {242003} (\bibinfo {year} {2012})},\ \Eprint
  {http://arxiv.org/abs/1112.4423} {arXiv:1112.4423 [hep-ph]} \BibitemShut
  {NoStop}%
\bibitem [{\citenamefont {Binosi}\ and\ \citenamefont
  {Theussl}(2004)}]{Binosi:2003yf}%
  \BibitemOpen
  \bibfield  {author} {\bibinfo {author} {\bibfnamefont {D.}~\bibnamefont
  {Binosi}}\ and\ \bibinfo {author} {\bibfnamefont {L.}~\bibnamefont
  {Theussl}},\ }\href {http://jaxodraw.sourceforge.net/} {\bibfield  {journal}
  {\bibinfo  {journal} {Comput. Phys. Commun.}\ }\textbf {\bibinfo {volume}
  {161}},\ \bibinfo {pages} {76} (\bibinfo {year} {2004})},\ \Eprint
  {http://arxiv.org/abs/hep-ph/0309015} {hep-ph/0309015} \BibitemShut {NoStop}%
\bibitem [{\citenamefont {Binosi}\ \emph {et~al.}(2009)\citenamefont {Binosi},
  \citenamefont {Collins}, \citenamefont {Kaufhold},\ and\ \citenamefont
  {Theussl}}]{Binosi:2008ig}%
  \BibitemOpen
  \bibfield  {author} {\bibinfo {author} {\bibfnamefont {D.}~\bibnamefont
  {Binosi}}, \bibinfo {author} {\bibfnamefont {J.}~\bibnamefont {Collins}},
  \bibinfo {author} {\bibfnamefont {C.}~\bibnamefont {Kaufhold}}, \ and\
  \bibinfo {author} {\bibfnamefont {L.}~\bibnamefont {Theussl}},\ }\href
  {\doibase doi:10.1016/j.cpc.2009.02.020} {\bibfield  {journal} {\bibinfo
  {journal} {Comput. Phys. Commun.}\ }\textbf {\bibinfo {volume} {180}},\
  \bibinfo {pages} {1709} (\bibinfo {year} {2009})},\ \Eprint
  {http://arxiv.org/abs/0811.4113} {arXiv:0811.4113 [hep-ph]} \BibitemShut
  {NoStop}%
\end{thebibliography}%

\end{document}